\documentclass[sigconf]{acmart}
\acmConference[arxiv]{arxiv pre-print versions}{2024}{}

\renewcommand\footnotetextcopyrightpermission[1]{}

\usepackage{amsmath, amsfonts}
\usepackage{algorithmic}
\usepackage{graphicx}
\usepackage{textcomp}
\usepackage{xcolor}
\usepackage{listings,listings-rust}
\usepackage{caption,subcaption}
\usepackage{enumerate}
\usepackage{multirow}
\usepackage{color}
\usepackage{colortbl}
\usepackage[ruled,vlined]{algorithm2e}
\usepackage{amsfonts}
\usepackage{threeparttable}
\usepackage{booktabs}
\usepackage{hyperref}

\definecolor{mygray}{gray}{.8}
\newtheorem{definition}{\textbf{Definition}}[section]
\newtheorem{prop}{\textbf{Property}}[section]

\def\BibTeX{{\rm B\kern-.05em{\sc i\kern-.025em b}\kern-.08em
    T\kern-.1667em\lower.7ex\hbox{E}\kern-.125emX}}
    
\begin{document}

\title{\textsc{rCanary}: Detecting Memory Leaks Across Semi-automated Memory Management Boundary in Rust}

\author{Mohan Cui}
\affiliation{%
  \institution{School of Computer Science, \\Fudan University}
  \country{}
}

\author{Hui Xu}
\authornote{Corresponding author.}
\affiliation{%
  \institution{School of Computer Science, \\Fudan University}
  \country{}
}

\author{Hongliang Tian}
\affiliation{%
  \institution{Ant Group}
  \country{}
}

\author{Yangfan Zhou}
\affiliation{%
  \institution{School of Computer Science, \\Fudan University}
  \country{}
}

\begin{abstract}
Rust is an effective system programming language that guarantees memory safety via compile-time verifications. It employs a novel ownership-based resource management model to facilitate automated deallocation. This model is anticipated to eliminate memory leaks. However, we observed that user intervention drives it into semi-automated memory management and makes it error-prone to cause leaks. In contrast to violating memory-safety guarantees restricted by the \textit{unsafe} keyword, the boundary of leaking memory is implicit, and the compiler would not emit any warnings for developers.

In this paper, we present \textsc{rCanary}, a static, non-intrusive, and fully automated model checker to detect leaks across the semi-automated boundary. We design an encoder to abstract data with heap allocation and formalize a refined leak-free memory model based on boolean satisfiability. It can generate SMT-Lib2 format constraints for Rust MIR and is implemented as a Cargo component. We evaluate \textsc{rCanary} by using flawed package benchmarks collected from the pull requests of open-source Rust projects. The results indicate that it is possible to recall all these defects with acceptable false positives. We further apply our tool to more than 1,200 real-world crates from crates.io and GitHub, identifying 19 crates having memory leaks. Our analyzer is also efficient, that costs 8.4 seconds per package.
\end{abstract}

\keywords{Memory Leak, Model Check, Semi-automated Memory Management, Boolean Satisﬁability, Ownership, Rust}

\maketitle

\section{Introduction}
Rust is an emerging programming language that focuses on memory safety and efficiency. Ownership-based resource management model (OBRM)~\cite{matsakis2014rust} can ensure memory safety but does not sacrifice performance via static analysis. It is a prominent design of zero-cost abstractions~\cite{jung2017rustbelt}. Specifically, Rust employs non-lexical lifetime (NLL) to deduce the minimum lifespan of each value\footnote{https://rust-lang.github.io/rfcs/2094-nll.html}. Thus, \textbf{automated deallocation} is achieved by inserting the \texttt{drop()} instruction when each value goes out of the life scope~\cite{klabnik2019rust}. As a system programming language, Rust enables users to evade automated deallocation and manage resources themselves, a lower control known as \textbf{semi-automated memory management}, to increase flexibility. Many companies and open-source projects thus switch to Rust, such as the web browser engine Servo~\cite{anderson2016engineering}, operating system Redox OS~\cite{light2015reenix,liang2021rustpi,redoxos}, LibOS Occlum~\cite{shen2020occlum,lankes2019exploring}, and the embedded system Tock OS~\cite{levy2015ownership,levy2017tock,levy2017multiprogramming}. In 2023, Rust has become the second programming language in the Linux kernel\footnote{https://lore.kernel.org/lkml/20210414184604.23473-1-ojeda@kernel.org/}.

We observed that the semi-automated memory management model may introduce memory leaks, such as issues listed in Table~\ref{table:RQ1}. Studying memory leaks~\cite{nagarakatte2012watchdog,hicks2004experience,balasubramanian2017system} is crucial for the system programming language. If the program consumes a significant quantity of memory but never releases it, the memory usage will continue to rise. Moreover, remote attackers can exploit memory leaks to launch a denial-of-service (DoS) attack~\cite{cweleak}. Despite numerous studies on leak detection for C/C++ programs~\cite{lattner2008llvm,cadar2008klee,jung2008practical}, to the best of our knowledge, no work has been conducted for Rust. This scope includes \textsc{FFIChecker}~\cite{li2022detecting} because it uses LLVM-IR to check for Rust-C FFI. In our latest empirical study on Rust security promise, memory leak emerged as a member in a categorization of safety requirements~\cite{cui2024unsafe}. Rust experts rank it fourth out of 19 categories in the vote for significance.

Leaking memory was not a safe operation from a temporal perspective. Until Rust version 1.0, the method of escaping from the ownership model was an unsafe intrinsic (\textit{i.e.,} \texttt{forget}~\cite{unsafeforget}). It was later marked as a safe method because it does not violate memory safety directly. However, developers may thus find it hard to pinpoint ownership leaks because this operation is not restricted to the unsafe scope, and the compiler does not detect it either. According to our statistics, semi-automated memory management is prevalently used in the Rust community. As of 2023-01-30, we mined the latest crates from crates.io~\cite{cratesio} and discovered that 5,717 repositories out of 103,516 employed this method. Compared to 21,506 crates using \textit{unsafe} keyword~\cite{astrauskas2020programmers,evans2020rust}, the potential vulnerability is a non-trivial problem for us to investigate.

We propose utilizing static analysis to detect leaks that cross the semi-automated boundary. Leak detection employs a taint-sink approach: the object created at the tainted site (construction) must reach the sink site (destruction)~\cite{sui2014detecting}. There are two types of existing static approaches: \textit{iterative data-flow analysis}~\cite{orlovich2006memory} and \textit{sparse value-flow analysis}~\cite{sui2016svf}. To our knowledge, the current work is ineffective for Rust. The former monitors the values at each program point throughout the control flow, which is flow-sensitive. However, it lacks support for Rust syntax, such as ownership movement. The latter tracks the values sparsely from the define-use chains or the single static assignment (SSA)~\cite{appel1998ssa}. It is more efficient but flow-insensitive. The sparse representation does not check arbitrary state-machine properties for all program paths. In addition, some work infers the owner for C/C++ programs based on boolean satisfiability~\cite{heine2003practical}, but it needs to be aligned with the ownership design.

This paper introduces \textsc{rCanary}, a static, non-intrusive, and fully automated model checker to detect leaks across the semi-automated boundary. It consists of an encoder to transform value based on its type and leverages a refined leak-free model over the ownership model. The encoder can convert the typed value into a bit vector to simulate construction. The element of the bit vector indicates which field contains heap allocation. The leak-free model is composed of formal rules based on boolean satisfiability to trace the deallocation. Our analyzer uses the encoder and formal rules to generate constraints for Rust MIR in SMT-Lib2 syntax~\cite{smtlib}, forming a 0–1 integer equality list. We implemented \textsc{rCanary} as an external component in \texttt{Cargo} and applied it to the existing flawed pull-request benchmarks collected from GitHub. The result demonstrates that it can recall all the issues that match our leak pattern and identify 19 crates with memory leaks in real-world Rust crates.

Our main contributions are summarized as follows:

\begin{enumerate}[0]
\item[$\bullet$] \noindent We studied the issues of memory leaks across the semi-automated memory management boundary in Rust and outlined two bug patterns. We developed algorithms to identify them: an encoder to abstract data having heap allocation and a leak-free memory model based on formal rules.
\item[$\bullet$] \noindent We implemented \textsc{rCanary}, a static, non-intrusive, and fully automated model checker that can detect leaks for Rust packages. It provides user-friendly diagnostics to pinpoint the buggy snippet. \textsc{rCanary} is open-sourced and can be modified for other research.
\item[$\bullet$] \noindent We evaluate the effectiveness, usability, and efficiency of \textsc{rCanary}. It can recall all bugs in the benchmark collected from GitHub and discover 19 crates with memory leaks in open-source Rust projects. It is also capable of executing fast scans for the Rust ecosystem.
\end{enumerate}

We first present the root causes of memory leaks in the ownership model and illustrate two bug patterns (\S\ref{sec:problem}). Then overview the system architecture of the analyzer because our algorithm has two critical components (\S\ref{sec:design}). We detail how to abstract data and construct values (rtoken) in our model, an encoder that generates bit vectors based on types (\S\ref{sec:layout}) and present a leak-free model, explaining how resources be used to detect memory leaks (\S\ref{sec:rules}).

\section{Problem Statement}\label{sec:problem}

\subsection{Why Leaks in the Ownership Model?}
The Rust ownership model is derived from the programming idiom Resource Acquisition Is Initialization (RAII)~\cite{ramananandro2012mechanized} used in C++,  where resources are acquired and released by objects. Specifically, the resource is acquired during the initialization and released during the destruction, ensuring proper resource management and preventing leaks. In the ownership model, when the user initializes a value, the compiler will calculate the program point to insert the \texttt{drop()} instruction and release the resource~\cite{matsakis2014rust}. This instruction typically binds to object variables that implement \texttt{Drop} trait~\cite{drop}, and the object is called \textbf{owner} of the resource (\textit{e.g.,} \texttt{Vec<T>}~\cite{vec}). Rust enables users not to instruct the automated deallocation that triggers semi-automated management. The escape hatch is \texttt{ManuallyDrop<T>}~\cite{manauallydrop}, a zero-cost wrapper that prevents the compiler from calling \texttt{T}'s destructor. Once the object is encapsulated in \texttt{ManuallyDrop}, it becomes an \textbf{orphan object}. As a smart pointer, \texttt{ManuallyDrop} can be dereferenced to disclose the generic field \texttt{\&T}, which can then be cast into the raw pointer \texttt{*T}. 

\begin{figure*}[t]
\begin{subfigure}[t]{0.49\textwidth}
	\includegraphics[width=\textwidth]{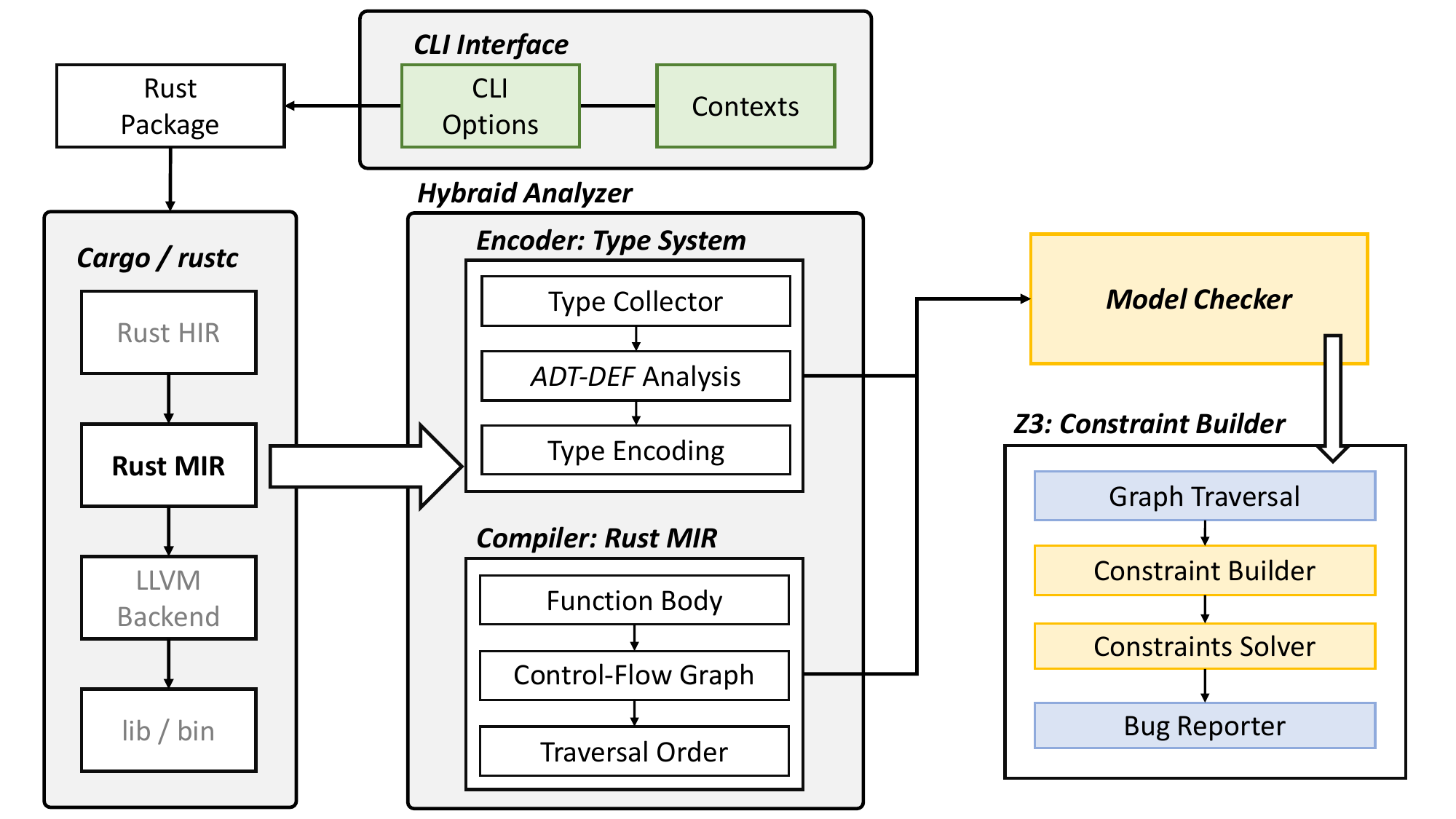}
	\caption{The holder of the rtoken for the heap item.}
	\label{fig:relation1}
\end{subfigure}
\begin{subfigure}[t]{0.49\textwidth}
	\includegraphics[width=\textwidth]{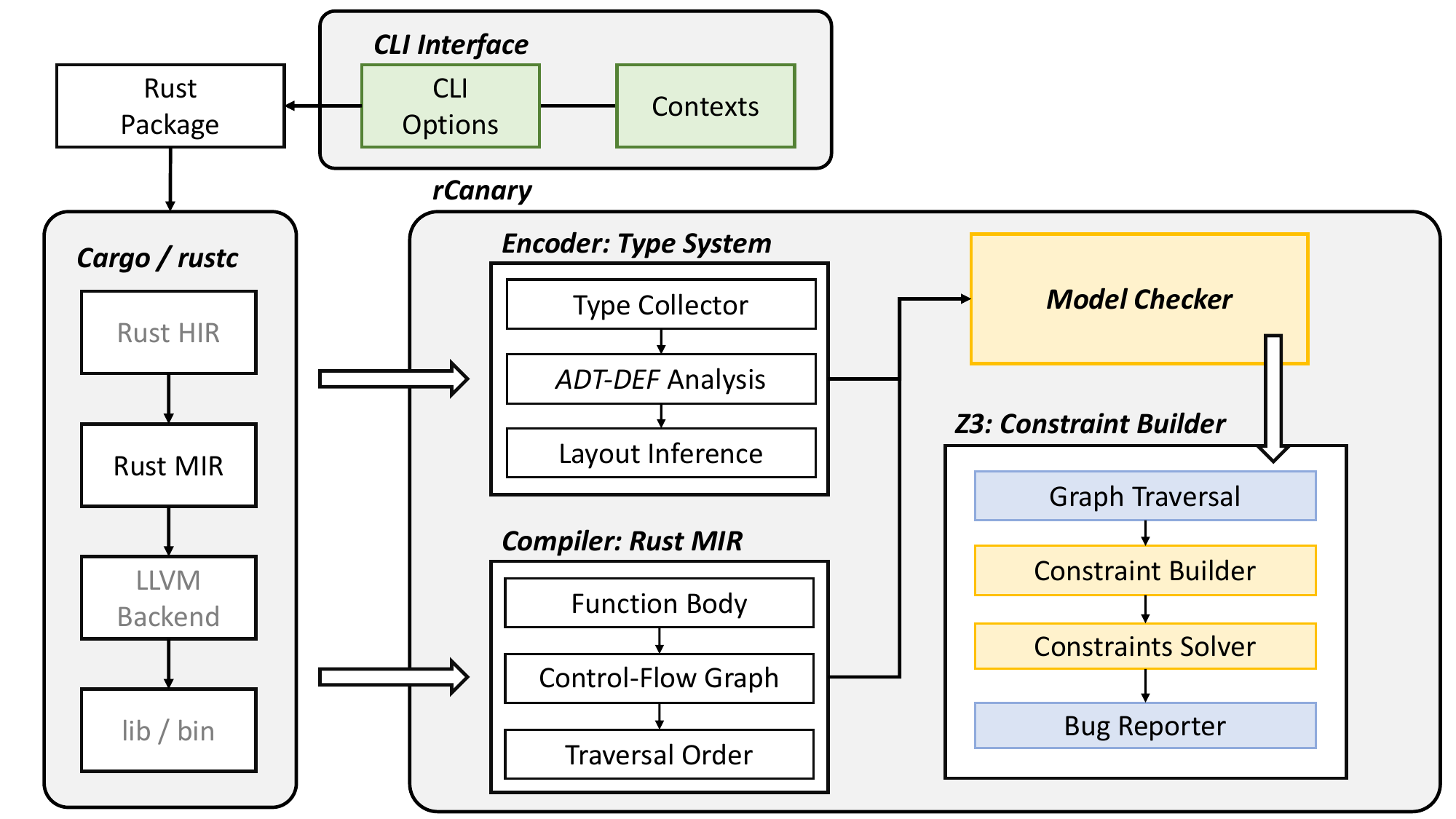}
	\caption{The assignments that cause the orphan-object and proxy-type leak.}
	\label{fig:relation2}
\end{subfigure}
\caption{The relationship between rtoken, Rust owner, and heap item. In Figure~\ref{fig:relation1}, the rtoken holders can be objects, references, and pointers. The stack frame and heap chunks reflect the real data storage. When an object exists, the object holds the rtoken (solid), while other pointers only point to it (dotted). If the object does not exist, pointer types hold the rtoken (solid). In Figure~\ref{fig:relation2}, the first argument of each method is wrapped with \texttt{ManuallyDrop} in the function body.}
\label{fig:relation}
\end{figure*}

\texttt{ManuallyDrop<T>} is a customizable design for toggling between automated deallocation and manual release. However, this design is vulnerable to leaks if the programmers neglect to release them. In today's Rust ecosystem, user intervention is a significant cause of leaks. In this paper, we summarize two typical leak patterns across semi-automated boundaries: (i) \textbf{orphan object} missing \texttt{drop()} instruction in Mid-level Intermediate Representation (MIR); (ii) \textbf{proxy type} missing manual field deallocation in its \texttt{Drop} implementation. They will be discussed in Sections~\ref{sec:oo} and~\ref{sec:pt}.

\begin{definition}[Heap Item, Rtoken]\label{def:heap_item}
	A heap item is an object containing the heap allocation with a fixed type (layout). It has an exclusive rtoken to monitor the deallocation, and the rtoken can be transferred among variables through the data flow.
\end{definition}

If one object escapes from the ownership model, we expect pointer types (including raw pointers \texttt{*} and references \texttt{\&}) to be responsible for deallocating them. As illustrated in Figure~\ref{fig:relation1}, this paper assumes that each value has an \textbf{RAII token (rtoken)} to monitor its deallocation and the definition is provided in Defition~\ref{def:heap_item}. If the object possesses the rtoken, its identity is equivalent to the owner. The primary distinction is that pointer types can hold the rtoken but can never be the owner.

\subsection{Motivating Example: Orphan Object}\label{sec:oo}
Creating an orphan object eliminates its \texttt{drop()} instruction in Rust MIR. Listing~\ref{list:me_wo} illustrates an example of an orphan-object leak, and we will give the definition below.

\begin{figure}[]
\begin{subfigure}[]{0.48\textwidth}
\begin{lstlisting}[language=Rust, style=colouredRust, label=list:me_wo, caption=Motivating example of orphan-object leak.\\]
fn main() {
    let mut buf = Box::new("buffer");(*@\label{me1:box_new}@*)
    // heap item 'buf' becomes an orphan object
    let ptr =
        &mut *ManuallyDrop::new(buf) as *mut _;(*@\label{me1:into_raw}@*)
    // leak by missing free operation on 'ptr'
+   unsafe { drop_in_place(ptr); }(*@\label{me1:fixed}@*)
}
\end{lstlisting}
\end{subfigure}

\begin{subfigure}[]{0.48\textwidth}
\begin{lstlisting}[language=Rust, style=colouredRust, label=list:me_pt, caption=Motivating example of proxy-type leak.\\]
struct Proxy<T> {
    ptr: *mut T,
}
impl<T> Drop for Proxy<T> {(*@\label{me2:define}@*)
    fn drop(&mut self) {
    // user should manually free the field 'ptr'
+      unsafe { drop_in_place(self.ptr); }(*@\label{me2:from_raw}@*)
    }
}
fn main() {
    let mut buf = Box::new("buffer");(*@\label{me2:box_new}@*)
    // heap item 'buf' becomes an orphan object
    let ptr = 
        &mut *ManuallyDrop::new(buf) as *mut _;(*@\label{me2:box_into_raw}@*)
    let proxy = Proxy { ptr };(*@\label{me2:proxy1}@*)
    // leak by missing free 'proxy.ptr' in drop
}
\end{lstlisting}
\end{subfigure}
\caption{Motivating examples of orphan-object and proxy-type issues detected in \textsc{rCanary}, caused by the lack of manual deallocation towards \texttt{ManuallyDrop} values.}
\label{fig:motivation}
\end{figure}

\begin{definition}[Orphan Object]\label{def:orphan_object}
	An orphan object is the heap item wrapped by the smart pointer \texttt{ManuallyDrop}.
\end{definition}

In Listing~\ref{list:me_wo}, we first construct a heap item \texttt{Box<\&str>} by using \texttt{Box::new} (line~\ref{me1:box_new}). The initial holder of its rtoken is \texttt{buf}. The method \texttt{ManuallyDrop::new} takes the heap item and encapsulates it with \texttt{ManuallyDrop} that creates an orphan object (line~\ref{me1:into_raw}). Then, we dereference it and cast it as a raw pointer. The rtoken is thus handed to \texttt{ptr}. From the ownership perspective, this line consumes the owner, restrains deallocation, and creates a raw pointer for further usage.

As depicted in Figure~\ref{fig:relation1}, variable \texttt{b} is an owner and holds a rtoken of the heap item. It can release the resource when its lifetime ends. The pointers \texttt{p1} and \texttt{r1} merely point to the heap item and would not free it. In contrast, the heap items pointed by \texttt{p2} and \texttt{r2} are orphan objects as they do not have owners. Thus, the poniters should act as the rtoken holder to release the resources. Figure~\ref{fig:relation2} summarizes the workflow: if the orphan object is assigned to another variable, it will be leaked if the programmer does not release the resource explicitly. The fixed snippet employs \texttt{drop\_in\_place}~\cite{dropinplace} to free the heap item (line~\ref{me1:fixed}), indicating that \texttt{ptr} is the holder to free the resource and delete the rtoken in deallocation.

\subsection{Motivating Example: Proxy Type}\label{sec:pt}
The proxy-type issue is caused by the unsound implementation of the \texttt{Drop} trait for the user-defined type, which is a mutated problem of the orphan-object leak. The proxy type is defined below.

\begin{definition}[Proxy Type]
	A proxy type is a compound type having at least one field that stores an orphan object.
\end{definition}

Listing~\ref{list:me_pt} shows a leak caused by the proxy type \texttt{Proxy<T>} (line~\ref{me2:define}). After creating an orphan object $\texttt{Box<\&str>}$, \texttt{ptr} is assigned to the field of a \texttt{Proxy<T>} object (line~\ref{me2:proxy1}). However, the default \texttt{Drop} implementation is ignorant that the field \texttt{ptr} ought to be deallocated. Consequently, it may trigger a memory leak whenever a \texttt{Proxy<T>} object is dropped. To solve this issue, we manually free the raw pointer by calling \texttt{drop\_in\_place} on the field (line~\ref{me2:from_raw}).

As depicted in Figure~\ref{fig:relation2}, if an orphan object is assigned to a field, although the compiler will append \texttt{drop()} to the proxy type, programmers are responsible for implementing a sound \texttt{Drop} method to release its fields explicitly.

These patterns can encompass primary scenarios caused by the lack of manual deallocation towards \texttt{ManuallyDrop} values, with the difference of having \texttt{drop()} instructions in Rust MIR. Because the place where a value is assigned can be variables and fields, even valid for globals. The limitation is that we do not support some complex data structures, such as storing orphan objects in static arrays or dynamic vectors.

\section{Design}\label{sec:design}
This section overviews the design goals and presents the system architecture of \textsc{rCanary}.

\subsection{Design Goals}
As with most existing static analysis analyzers, \textsc{rCanary} analyzes a source code representation~\cite{cui2023safedrop,bae2021rudra,li2021mirchecker}. We propose combining data-flow analysis and boolean satisfiability to design a static, non-intrusive, and fully automated model checker on top of the Rust MIR. We have three design goals:

\begin{enumerate}[0]
\item[$\bullet$] \noindent \textbf{Abstraction.} \textsc{rCanary} focuses on heap values. It needs to recognize the values with heap allocation and encode them in a proper format for model check. The method should support the primary features of Rust types, such as generics, mono-morphizations, etc. We chose the MIR variable (a.k.a. local) as the target because it retains type information and passes type checks.
\item[$\bullet$] \noindent \textbf{Usability.} \textsc{rCanary} is designed as a general static analyzer for Rust projects. It should analyze each crate within a Rust package iteratively and report potential leaks across the semi-automated boundary. As an external component of the Rust toolchain, it should support \texttt{Cargo} projects and provide a user-friendly interface. The workflow should be fully automated, and no user annotation is required.
\item[$\bullet$] \noindent \textbf{Efficiency.} \textsc{rCanary} is based on the theory of boolean satisfiability. It traverses MIR functions to generate constraints and employs SMT solver Z3~\cite{moura2008z3} to solve those constraints. The computation overhead should be seriously considered. It should balance precision and efficiency to achieve a fast scan for the large-scale project, even for the Rust ecosystem.
\end{enumerate}

\subsection{System Architecture}\label{sec: system}
\begin{figure}
\centering
\includegraphics[width=0.49\textwidth]{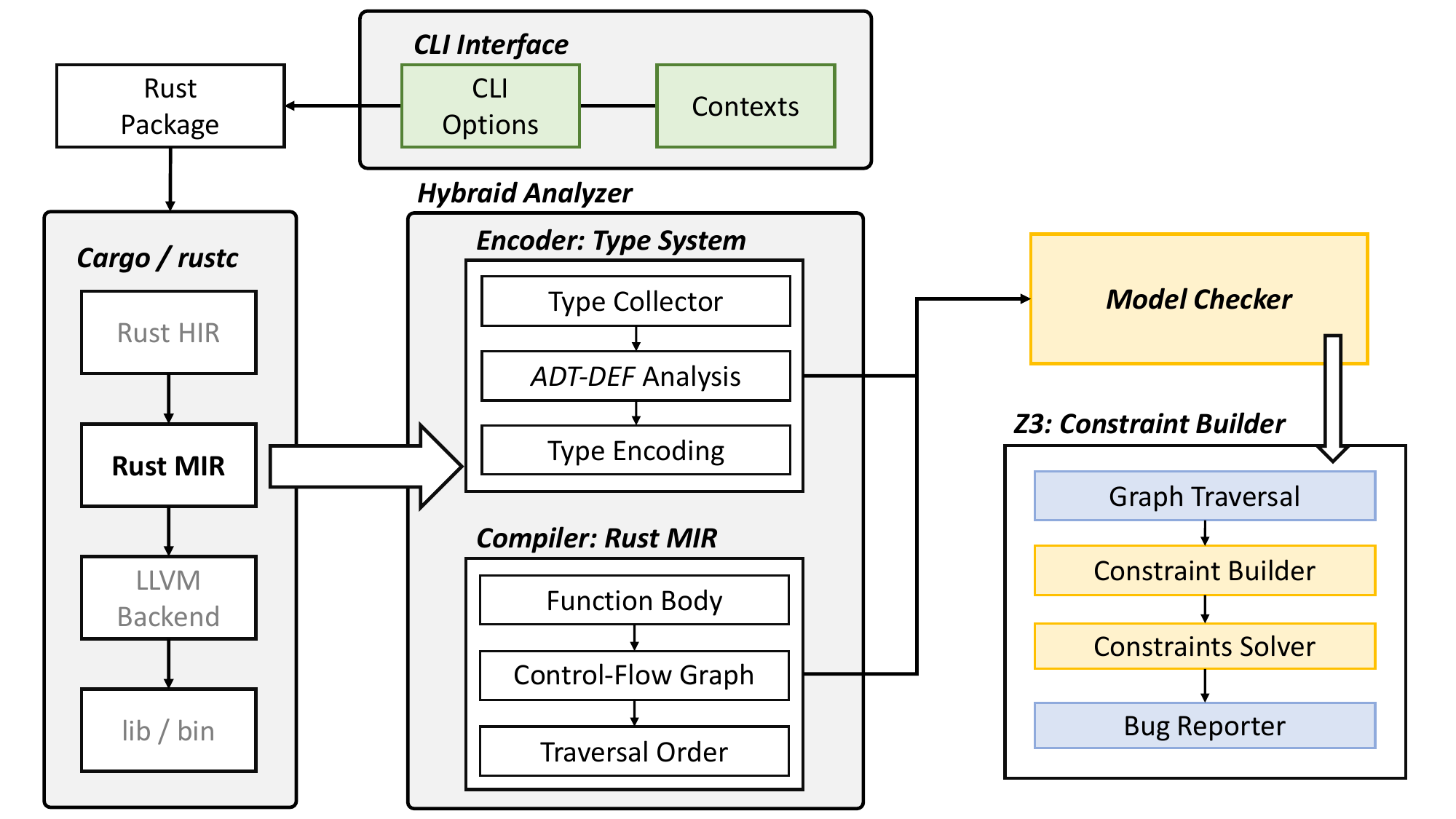}
\caption{The system architecture of \textsc{rCanary}.}
\label{fig:framework}
\end{figure}

The framework of \textsc{rCanary} consists of three components: the CLI interface, the hybrid analyzer, and the model checker, as shown in Figure~\ref{fig:framework}.

\subsubsection{CLI Interface} The front end is a CLI program installed in the toolchain. We implemented \textsc{rCanary} as an external component that can be invoked as a subcommand for \texttt{Cargo}. We provide secondary configurations to enhance customization but not impede compilation settings. It will be invoked after parsing the CLI parameters and initializing the configurations. The compilation metadata will be emitted after all MIR passes are finished. After analysis, the bug reporter will notify users of potential leaks with bug locations.

\subsubsection{Hybrid Analyzer} The first component of the backend is a hybrid analyzer developed on top of Rust MIR. It evaluates the node order of the control-flow graph and employs the Tarjan algorithm~\cite{tarjan} to eliminate loops. To model heap values, we designed an encoder that can transform arbitrary types into fixed-length bit vectors. It is efficient with linear complexity. It first analyzes the type definitions and then uses the result to accelerate encoding for mono-morphizations.

\subsubsection{Model Checker} The second part is a model checker. We design a leak-free memory model consisting of formal rules to detect leaks, which are flow-sensitive and field-sensitive. The model checker uses the hybrid analyzer to get the encoder and metadata of Rust MIR. It can encode values to generate the Z3 variables and apply formal rules for statements to generate the Z3 assertions. Finally, the solver will be activated to solve constraints and the bug reporter will emit issues with code spans for developers.

\section{Encoder: Data Abstraction}\label{sec:layout}
This section introduces the approach to constructing rtoken. It is an encoder that depicts data abstraction based on Rust types. The encoder can identify heap items and transform them into bit vectors.
 
\subsection{Data Abstraction}\label{sec:layout_abstraction}

\begin{table}\small
\caption{The Rust type classification in the encoder.}
\label{table:type_classfy}
\centering
\resizebox{\linewidth}{!}{
\begin{threeparttable}
\begin{tabular}{cc>{\columncolor{mygray}}ccclc}%
	\toprule[1pt]
	\multirow{2}{*}{\textbf{Classification}}  & \multirow{2}{*}{\textbf{Type}}   & \multicolumn{3}{c}{\multirow{1}{*}{\textbf{Encoding Representation}}}     & \multicolumn{1}{c}{\multirow{2}{*}{\textbf{Example}}} \\
	\cmidrule(r){3-5}
	& & \textbf{Default} & \textbf{Length} & \textbf{Data Flow} & \\
	\midrule[1pt]
	\multirow{3}{*}{\textbf{Primitive}}    & \textbf{Boolean}      & [0]       & 1       & -             & \texttt{true}, \texttt{false}\\
	                                       & \textbf{Numeric}      & [0]       & 1       & -             & \texttt{i8}, \texttt{u16}, \texttt{isize}\\
	                                       & \textbf{Texual}       & [0]       & 1       & -             & \texttt{char}, \texttt{str}\\
	\midrule[1pt]
	\multirow{7}{*}{\textbf{Object}}       & \textbf{Array}        & E         & 1       & Ctor / Ctx    & \texttt{[i32;1]}\\
	                                       & \textbf{Struct}       & E         & Fixed   & Ctor / Ctx    & \texttt{Vec<T>}\\
	                                       & \textbf{Enumerate}    & E         & Fixed   & Ctor / Ctx    & \texttt{Option<T>}\\
	                                       & \textbf{Union}        & E         & Fixed   & Ctor / Ctx    & \texttt{MaybeUninit<T>}\\
	                                       & \textbf{Tuple}        & E         & Fixed   & Ctor / Ctx    & \texttt{(i32, String)}\\
	                                       & \textbf{Trait Object} & [1]       & 1       & Ctor / Ctx    & \texttt{dyn Clone+Copy}\\
	                                       & \textbf{Parameter}    & [1]       & 1       & Definition    & \texttt{T}, \texttt{A}, \texttt{S}\\
	\midrule[1pt]
	\multirow{3}{*}{\textbf{Pointer}}      & \textbf{Slice}        & [0]       & Fixed   & Ctx           & \texttt{\&str}, \texttt{\&[i32;1]}\\
	                                       & \textbf{Reference}    & [0]       & Fixed   & Ctx           & \texttt{\& \_}, \texttt{\&mut \_}\\
	                                       & \textbf{Raw Pointer}  & [0]       & Fixed   & Ctx           & \texttt{*const \_}, \texttt{*mut \_}\\
	\bottomrule[1pt]
\end{tabular}
\begin{tablenotes}
	\footnotesize
	\item[1] Abbreviation: E: Type Encoding, Ctor: Constructor, Ctx: Context.
	\item[2] The classification is from Rust reference but partially modified in this paper.
	\item[3] The parameter types (generics) are defined in signatures with no constructors.
	\item[4] Zero-sized types (ZSTs) are object types not listed in this table. Our encoder regards ZSTs as zeroed-length vectors ([ ]), and the operations are optimized as NOP in Z3.
\end{tablenotes}
\end{threeparttable}
}
\end{table}

A Rust project contains a set of types among all its variables. As shown in Table~\ref{table:type_classfy}, the Rust types can be classified into primitives, objects, and pointers~\cite{reference}. The object types have a set of constructors and a unique destructor. The constructor can be a function with a list of arguments (\textit{e.g.,} \texttt{Vec::new}~\cite{vecnew}) or the local initialization. The destructor is the unique \texttt{Drop} implementation associated with this type.

Each type has a rtoken constructor to illustrate the data abstraction in the fully initialized state. The rtoken is represented as a fixed-length bit vector. Its member can take on values $\sigma\in\{-1, 0, 1\}$. These elements become the lifted boolean values~\cite{moura2008z3}, where $\sigma=-1$ is an uninitialized value, $\sigma=1$ is holding a heap item, and $\sigma=0$ is the opposite (including not holding a heap item, stack-allocated value, freed value or moved value). The lifted boolean values form a semi-lattice~\cite{moller2012static}: the top element $\top$ is $-1$, and the bottom element $\bot$ is $0$. We separate Rust types into enumeration types (\textit{e.g.,} \texttt{IpAddr}~\cite{ipaddr}) and others (\textit{e.g.,} \texttt{Vec}~\cite{vec}), because different variants in enum types can carry different types of data~\cite{reference}. Thus, we define the rtoken as below:

\begin{definition}[Rtoken]\label{def:ctor}
	For the non-enumeration type, rtoken is a fixed-length bit vector. Each member denotes whether this initialized field is a heap item. For the enumeration type, if its variance can be determined, rtoken is a fixed bit vector of its associated data. Otherwise, the extent is zeroed (unknown).
\end{definition}

 This data abstraction is field-sensitive as it uses the member inside the bit vector to simulate the field. Notably, it disregards the alignment and focuses on the field index; namely, the index corresponds to the field order specified in the type definition. Enumeration types require a variance index (\textit{i.e.,} discriminant) as the length of the rtoken generated by different variants may be different. If the discriminant is lost, the rtoken is unknown. This rule can be applied to compound types that have enumeration fields.

\subsection{ADT-Definition Analysis}
ADT-Definition~\cite{adtdef} (AdtDef) analysis determines the possibility of a defined type becoming the heap item when fully initialized. The algorithm is in Algorithm~\ref{algo:layout}.

\begin{algorithm}[t]
\caption{\textsc{Encoder}: Construct the rtoken for the given type in the fully-initialized state, represented as a fixed-length bit vector.}
\label{algo:layout}
\LinesNumbered

\SetKwFunction{AdtDefAnalysis}{AdtDefAnalysis}
\SetKwFunction{EncodeField}{EncodeField}
\SetKwFunction{CollectTypes}{CollectTypes}
\SetKwFunction{ExtractAdtDefs}{ExtractAdtDefs}
\SetKwFunction{AnalyzeDependencies}{AnalyzeDependencies}
\SetKwFunction{InverseToposort}{InverseToposort}
\SetKwFunction{HeapItemUnit}{hasHeapItemUnit}
\SetKwFunction{IsolatedParameter}{hasIsolatedParameter}
\SetKwFunction{IsIsolatedParameter}{isIsolatedParameter}

\SetKwProg{Fn}{Function}{}{end}

\BlankLine	

\KwIn{\textit{mir}: mir body of given function}
\KwOut{\textit{cache}: cache of AdtDef analysis results}

\BlankLine

\Fn{\AdtDefAnalysis{mir}}{                                                \label{fn1}
\textit{tys} $\leftarrow$ \CollectTypes{mir}                              \label{collector}\\
\textit{adtdefs} $\leftarrow$ \ExtractAdtDefs{tys}                        \label{extract}\\
\textit{depG} $\leftarrow$ \AnalyzeDependencies{adtdefs}                     \label{denpendency}\\
\textit{cache} $\leftarrow$ $\emptyset$\\
\ForEach {depCC in depG}{
	\textit{depQueue} $\leftarrow$ \InverseToposort{depCC}                 \label{order}\\
	\ForEach {adtdef in depQueue}{
	\textit{result.heap\_item} $\leftarrow$ \HeapItemUnit{adtdef}          \label{judge1}\\
	\textit{result.isolated\_parameter} $\leftarrow$ \IsolatedParameter{adtdef}     \label{judge2}\\
	\textit{cache} $\leftarrow$ \textit{result}                        \label{cache}\\
	}
}
\Return \textit{cache}\\
}

\BlankLine

\KwIn{\textit{ty}: type metadata in MIR}
\KwOut{\textit{bool}: the result of field encoding}

\Fn{\EncodeField{ty}}{                                         \label{fn2}
\If{\HeapItemUnit{ty.adtdef}}{                                 \label{isot}
	\Return \textit{TRUE}\\
}
\ForEach {generic in ty.genericargs}{
	\If{\IsIsolatedParameter{generic}}{          \label{isrp}
		\Return \EncodeField{generic}\\
	}
}
\Return \textit{FALSE} \\
}

\end{algorithm}

\noindent\textbf{Preprocessing.} An arbitrary type consists of a type definition (AdtDef~\cite{adtdef}) and a list of the generic arguments (GenericArgs~\cite{genericargs}). For instance, \texttt{Vec<i32>} consists of the AdtDef \texttt{Vec<\_>} and the generic argument \texttt{i32}. In preprocessing, we invoke an inter-procedural visitor \texttt{CollectTypes} to collect all encountered types into the set \textit{adtdefs} (line~\ref{collector}). Those types are located not only in the local crate but also in dependencies. Then we traverse each type and extract the definition of compound types by using \texttt{ExtractAdtDefs} (line~\ref{extract}). Finally, fields are flattened, so we can generate a dependency graph of AdtDefs \textit{depG} in \texttt{AnalyzeDependencies} (line~\ref{denpendency}). Since Rust restricts recursion in compound types, it still permits recursion in GenericArgs (\textit{e.g.,} \texttt{Vec<Vec<i32>\>>}). Thus, we break the recursion within GenericArgs and construct a directed acyclic graph (DAG).

\noindent\textbf{AdtDef and GenericArgs.} We will analyze if each type is a heap item or is able to become one in mono-morphizations. It is crucial to define the form of the heap item and tackle the generics. The heap-item unit in Definition~\ref{def:unit} is the identity of heap items derived from the collection types in the standard library. \texttt{PhantomData}~\cite{phantomdata} is the marker that informs the compiler that one type is a container and stores the typed value on the heap. If it is held by a flattened type, at least one field is allocated on the heap. In Definition~\ref{def:isolated}, the isolated parameter demonstrates whether a generic parameter can be further mono-morphized as a heap item. If an isolated parameter mono-morphizes into a heap-item type, this mono-morphization will generate a heap item.

\begin{definition}[Heap-item Unit]\label{def:unit}
	A heap-item unit has a \texttt{PhantomData<T>} field with a typed generic parameter and a pointer field to store the value \texttt{T}.
\end{definition}

\begin{definition}[Isolated Parameter]\label{def:isolated}
	An isolated parameter is a stand-alone and typed generic parameter (\textit{e.g.,} \texttt{T} but not \texttt{\&T}).
\end{definition}

\noindent\textbf{Result Solving.} For each connected component \textit{depCC}, we use \texttt{InverseToposort} to compute the subgraph's inverse topological order (line~\ref{order}). We will recursively search whether each AdtDef contains the heap-item unit and the isolated parameter (lines~\ref{judge1},~\ref{judge2}). The result comprises two parts: (i) a boolean value indicating if it contains a heap item; (ii) a boolean vector indicating if each generic parameter is an isolated parameter (\textit{e.g.,} the result of the vector \texttt{Vec<T,A>}~\cite{vec} is $(1)$ $[0,1]$, as shown in Figure~\ref{fig:encoding}). It can be simplified by using inverse topo-order. We first determine whether each field is a heap-item unit or an isolated parameter. If neither, we can read from the successive node to identify if this field contains a heap-item unit or isolated parameter directly. Furthermore, the results are cached as one AdtDef being analyzed (line~\ref{cache}).

\subsection{Type Encoding: Rtoken Constructor}
The encoder constructs the rtoken as a bit vector by iteratively calling function \texttt{EncodeField} (line~\ref{fn2}) on each field to decide whether this field is a heap item in the fully initialized state. As the type of the field consists of AdtDef and GenericArgs, the encoder first queries the cache whether the AdtDef has a heap item unit (line~\ref{isot}). If not, it will search for each isolated parameter available and verify if its GenericArg projects to a heap-item type recursively (line~\ref{isrp}).

\begin{figure}[]
\includegraphics[width=0.5\textwidth]{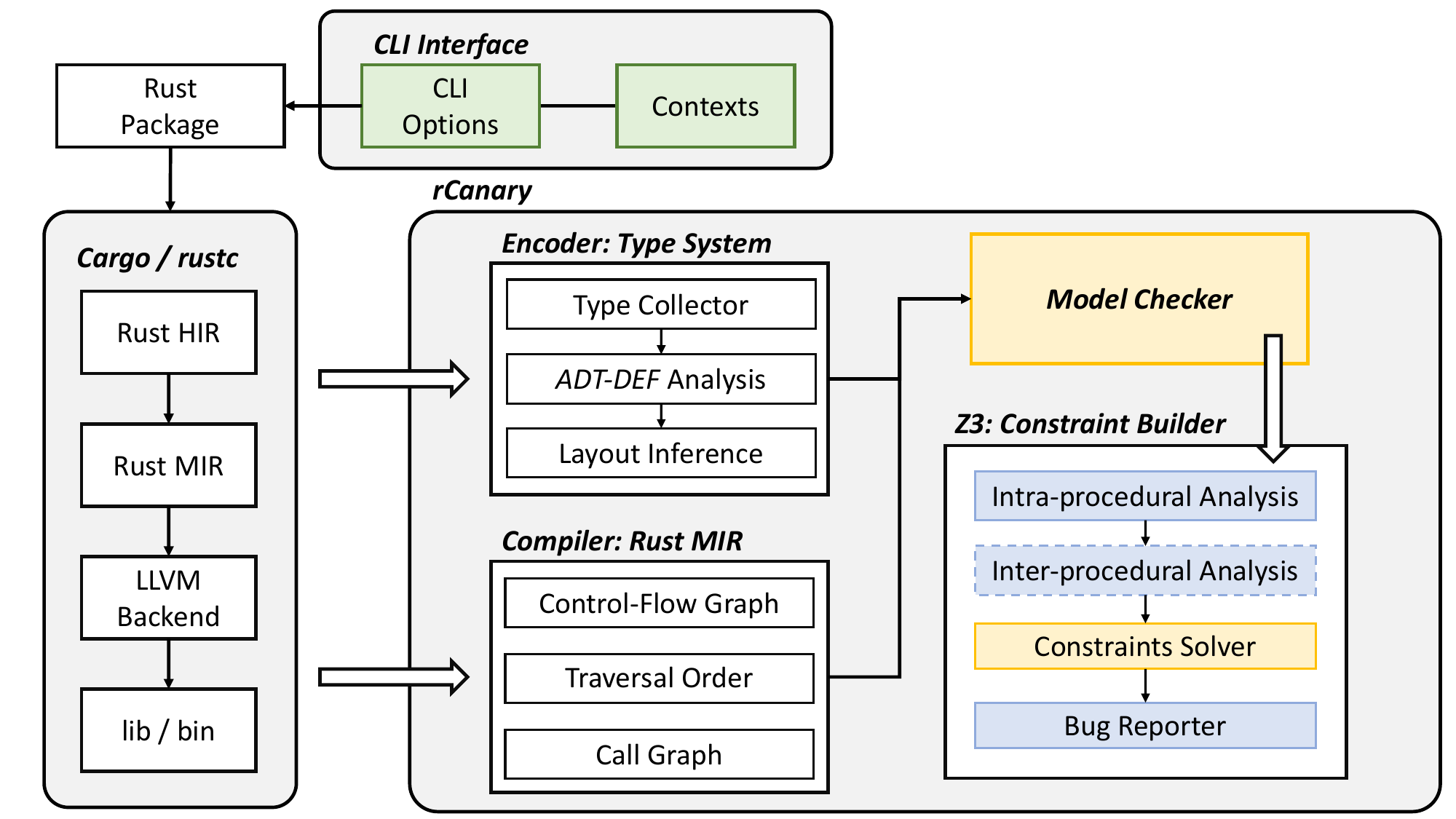}
\caption{Example of the encoder, including AdtDef analysis for \texttt{Vec<T,A>} and type encoding for \texttt{String}. The types have been flattened, forming a directed acyclic graph (DAG). Results are propagated forward using reverse topological sorting.}
\label{fig:encoding}
\end{figure}

We present an example in Figure~\ref{fig:encoding}. The rtoken constructor of \texttt{String}~\cite{string} is $[1]$. Our algorithm analyzes its field and finds it is a mono-morphization on \texttt{Vec}. Since \texttt{Vec<T, A>} contains a heap-item unit, the result of \texttt{String} is $[1]$ with no need to handle the generics. Likewise, the rtoken constructor of $\texttt{Vec<u8,Global>}$ is $[1,0]$. Our algorithm first analyzes the field \texttt{RawVec} and finds that it contains a heap-item unit. It then analyzes the second field \texttt{usize} and concludes the result of the second field is $0$.

This algorithm is efficient, and the maximal time complexity is $O(fd)$, where $f$ is the maximum size among fields and $d$ is the maximum depth among GenericArgs. AdtDef analysis accelerates encoding via a cached hash table. It is exempt from searching in fields recursively or suffering from various mono-morphizations. Trait objects and parameters are conservatively regarded as heap items, as shown in Table~\ref{table:type_classfy}. At last, the rtoken constructors of the pointer types are always $\vec{0}$, and they should be inferred from the context as explained in Section~\ref{sec:rules}.

\begin{figure*}
\centering

$^{^{[MERGE\ \phi]}}\quad\frac{M\ \vdash\ \forall x:\vec{\sigma_{x}} \quad \vec{\sigma_{x}}^{\prime}\ \mathbf{new} \quad C^{\prime}=C \wedge\left\{\vec{\sigma_{x}}^{\prime}=\cap_i\right\}}{H ;\ M ;\ C\ \vdash\ \phi(x)\ \Rightarrow\ M\left[x \mapsto \vec{\sigma_{x}}^{\prime}\right] ;\ C^{\prime}}\qquad$
$^{^{[RETURN]}}\quad\frac{M\ \vdash\ \forall x:\vec{\sigma_{x}} \quad C^{\prime}=C \wedge\left\{\vec{\sigma_{x}}=\vec{0} \right\}}{H ;\ M ;\ C\ \vdash\ \mathbf{return}\ \Rightarrow\ C^{\prime}}$

$^{^{[ASGNM]}}\quad{\frac{M \ \vdash \ x:\vec{\sigma_{x}},\ y:\vec{\sigma_{y}} \quad \vec{\sigma_{x}}',\ \vec{\sigma_{y}}' \ \mathbf{new} \quad C'=C\land\{ \vec{\sigma_{y}}=\vec{0} \} \land \{ \vec{\sigma_{x}}' =\vec{0}\} \land \{ \vec{\sigma_{y}}' = \vec{\sigma_{x}} \}} {H;\ M;\ C \ \vdash\ y= \mathbf{move} \ x \  \Rightarrow \  M[y \mapsto \vec{\sigma_{y}}',\ x \mapsto \vec{\sigma_{x}}'];\ C' }}$

$^{^{[ASGNM-READF]}}\quad{\frac{M \ \vdash \ x.f:\sigma_{f}, \ y:\vec{\sigma_{y}} \quad \sigma_{f}', \ \vec{\sigma_{y}}' \ \mathbf{new} \quad C'=C\land\{ \vec{\sigma_{y}}=\vec{0} \} \land \{\sigma_{f}' = 0\} \land \{ \vec{\sigma_{y}}' = \mathtt{Ext}(\sigma_{f})\}}{H;\ M;\ C \ \vdash \ y=\mathbf{move} \ x.f \ \Rightarrow \ M[y \mapsto \sigma_{y}', \ x.f \mapsto \sigma_{f}'];\ C'}}$

$^{^{[ASGNM-WRITEF]}}\quad{\frac{M \ \vdash \ x:\vec{\sigma_{x}}, \ y.f:\sigma_{f} \quad \vec{\sigma_{x}}', \ \sigma_{f}' \ \mathbf{new} \quad C'=C\land\{ \sigma_{f}=0 \} \land \{\vec{\sigma_{x}}'=\vec{0}\} \land \{\sigma_{f} = \mathtt{Srk}(\vec{\sigma_{x}})\}}{H; \ M; \ C \ \vdash \ y.f= \mathbf{move}\ x \ \Rightarrow \ M[y.f \mapsto \sigma_{f}',\ x \mapsto \vec{\sigma_{x}}'];\ C'}}$

$^{^{[ASGNM-FTOF]}}\quad{\frac{M \ \vdash \ x.f:\sigma_{x},\ y.f:\sigma_{y} \quad \sigma_{x}',\ \sigma_{y}' \ \mathbf{new} \quad C'=C\land\{ \sigma_{y}=0 \} \land \{ \sigma_{x}'=0\} \land \{ \sigma_{y}' = \sigma_{x} \}} {H;\ M;\ C \ \vdash\ y.f= \mathbf{move} \ x.f \ \Rightarrow \  M[y.f \mapsto \sigma_{y}',\ x.f \mapsto \sigma_{x}'];\ C' }}$

$^{^{[ASGNC]}}\quad{\frac{M \ \vdash \ x:\vec{\sigma_{x}},\ y:\vec{\sigma_{y}} \quad \vec{\sigma_{x}}',\ \vec{\sigma_{y}}' \ \mathbf{new} \quad C'=C\land\{ \vec{\sigma_{y}}=\vec{0} \} \land \{ \{\vec{\sigma_{x}}' = \vec{\sigma_{x}} \land \vec{\sigma_{y}}'=\vec{0}\} \lor \{\vec{\sigma_{y}}' = \vec{\sigma_{x}}\land\vec{\sigma_{x}}'=\vec{0}\}\}}{H;\ M;\ C \ \vdash \ y=x \ \Rightarrow \ M[y \mapsto \vec{\sigma_{y}}',\ x \mapsto \vec{\sigma_{x}}'];\ C'}}$

$^{^{[ASGNC-READF]}}\quad{\frac{M \ \vdash \ x.f:\sigma_{f}, \ y:\vec{\sigma_{y}} \quad \sigma_{f}', \ \vec{\sigma_{y}}' \ \mathbf{new} \quad C'=C\land\{ \vec{\sigma_{y}}=\vec{0} \} \land \{ \{\sigma_{f}' = \sigma_{f} \land \vec{\sigma_{y}}' = \vec{0} \} \lor \{\vec{\sigma_{y}}' = \mathtt{Ext}(\sigma_{f}) \land \sigma_{f}'=0 \}\}}{H;\ M;\ C \ \vdash \ y= \ x.f \ \Rightarrow \ M[y \mapsto \sigma_{y}',\ x.f \mapsto \sigma_{f}'];\ C'}}$

$^{^{[ASGNC-WRITEF]}}\quad{\frac{M \ \vdash \ x:\vec{\sigma_{x}}, \ y:\sigma_{f} \quad \vec{\sigma_{x}}', \ \sigma_{f}' \ \mathbf{new} \quad C'=C\land\{ \sigma_{f}=0 \} \land \{ \{\vec{\sigma_{x}}' = \vec{\sigma_{x}} \land \sigma_{f}' = 0 \} \lor \{\sigma_{f}' = \mathtt{Srk}(\vec{\sigma_{x}}) \land \vec{\sigma_{x}}'=\vec{0} \}\}}{H;\ M;\ C \ \vdash \ y.f=x \ \Rightarrow \ M[y.f \mapsto \sigma_{f}',\ x \mapsto \vec{\sigma_{x}}'];\ C'}}$

$^{^{[ASGNC-FTOF]}}\quad{\frac{M \ \vdash \ x.f:\sigma_{x},\ y.f:\sigma_{y} \quad \sigma_{x}',\ \sigma_{y}' \ \mathbf{new} \quad C'=C\land\{ \sigma_{y}=0 \} \land \{ \{\sigma_{x}' = \sigma_{x} \land \sigma_{y}'=0\} \lor \{\sigma_{y}' = \sigma_{x} \land \sigma_{x}'=0\}\}}{H;\ M;\ C \ \vdash \ y.f=x.f \ \Rightarrow \ M[y.f \mapsto \sigma_{y}',\ x.f \mapsto \sigma_{x}'];\ C'}}$

\caption{The intra-procedural rules in the leak-free memory model.}
\label{fig:constraint}
\end{figure*}

\section{Leak-free Model: Formal Rules}\label{sec:rules}
In this section, we present a refined memory model consisting of formal rules that can be applied to generate constraints on top of Rust MIR to detect leaks across the semi-automated management boundary.
 
\subsection{Leak-free Typed Memory Model}
Each heap item has a unique rtoken to monitor the deallocation as soon as it is constructed. The refined leak-free model has the form:

\begin{center}
	$H;\ M;\ C \ \vdash \ s \ \Rightarrow \ M';\ C'$
\end{center}

$H$ is the set of the heap items bounded with Rust types, $M$ is a mapping of holder candidates for each rtoken in each program point, and $C$ is a set of constraints. After analyzing a statement, this model will update the mapping $M'$ and the constraint set $C'$. Our leak-free model is monotonic~\cite{nielson2004principles} as the bit vector becomes a semi-lattice~\cite{moller2012static}. In this model, a leak-free function should satisfy the following properties:

\begin{prop}\label{prop:1}
	There exists one and only one rtoken for each heap item allocated but not deallocated.
\end{prop}

\begin{prop}\label{prop:2}
	Every heap item except the return value must execute \texttt{drop()} to free the rtoken before return.
\end{prop}

\begin{prop}\label{prop:3}
	The \texttt{drop()} instruction must free all necessary fields.
\end{prop}

\begin{prop}\label{prop:4}
	The type associated with the rtoken is fixed across statements and function calls.
\end{prop}

Property~\ref{prop:1} ensures that the rtoken is exclusive at each program point. Like Rust owner, rtoken can be transferred through statements and functions, but the holder is extended to pointer types (\textit{i.e.,} references and raw pointers). Property~\ref{prop:2} indicates that each heap item must execute its destructor at the end of its lifetime to release the resource, but it cannot guarantee that the deallocation method will free all required fields. This deficiency is remedied by Property~\ref{prop:3}, which improves the soundness of the \texttt{Drop} implementation. Noncompliance with Property~\ref{prop:2} is the primary cause of orphan-object leaks, while Property~\ref{prop:3} targets proxy-type issues.

Property~\ref{prop:4} is an additional requirement. Table~\ref{table:syntax} lists the core syntax of Rust MIR for leak detection~\cite{jung2017rustbelt,reed2015patina,wang2018krust}. As in Table~\ref{table:syntax}, addressing has two distinct forms, \texttt{Reference} and \texttt{AddressOf}, which create a reference and a raw pointer, respectively. The syntax for dereferencing is in \texttt{Place} item. The pointer types represent the types of pointed places (\textit{e.g.,} \texttt{\&Vec<u8>}) but can be cast easily. This property aims at restricting the ambiguity among pointee places. Since rtokens are constructed by the typed encoder, the type held by pointers is fixed in $H$ and can be inferred through contexts. The default rtoken constructor for pointer types is [0], so pointers cannot be the holder without contexts. Property~\ref{prop:4} simplifies pointer access by binding each variable with a fixed type. Thus, we do not need to distinguish objects and pointers (\textit{i.e.,} addressing and dereferencing).

\subsection{Assignment Rules}

\subsubsection{Graph Traversal} Functions are represented as control flow graphs in MIR, which are directed graphs composed of basic blocks as nodes and control flows as edges. \textsc{rCanary} will extract each function in the project, excluding dependencies, as the entry to start the analysis. We employ the Tarjan algorithm~\cite{tarjan} to separate each strongly connected component (SCC) into DAG, then calculate the topological order for each DAG. In each node, we will analyze the statements from the top-down order, which is flow-sensitive. Once it encounters a node with multiple predecessors, it will merge the states of all preceding nodes using $\phi$ operation~\cite{moller2012static}. Since the $\phi$ function is monotonic, the algorithm will traverse loops for one time; namely, each node generates constraints only once.  As it is not an iterative dataflow analysis approach, we do not need to reach the fixed point in iterations; instead, we only focus on the transferring of rtokens. Once all the nodes have been analyzed, the generation phase terminates. So we obtain a set of constraints, that are SMT expressions and will be solved by Z3 solver.

\subsubsection{Syntactic Rules} All the assignments connect \texttt{Place} with \texttt{Rvalue} in Rust MIR. \texttt{move} (explicit) and \texttt{copy} (implicit) are different semantics~\cite{reference} with distinct rules. Due to space constraints, we only describe two representative rules.

\noindent \textbf{\texttt{move} Syntax.} The rule for \texttt{move} syntax is ${ASGNM}$. $\vec{\sigma_x}, \vec{\sigma_y}$ represent the prior states of whether variables $x$ and $y$ hold the rtoken, which are stored in $M$. $\vec{\sigma_x}', \vec{\sigma_y}'$ are the posterior states (\textbf{new}); term \textbf{new} represents the new states after executing the statement and these states do not exist in $M$ until the constraint building of this statement is finished. If the rvalue carries the rtoken, the holder should be explicitly switched to the lvalue. \texttt{move} indicates that the rvalue $x$ is no longer the holder. Thus, the constraint consists of three parts: (i) $\vec{\sigma_{y}}=\vec{0}$: since it will overwrite $y$, $\vec{\sigma_y}$ cannot hold the rtoken to prevent leaks, which satisfies Property~\ref{prop:2}; (ii) $\vec{\sigma_{x}}' =\vec{0}$: due to \texttt{move} syntax, the new state $\vec{\sigma_x}'$ is not the holder; (iii) $\vec{\sigma_{y}}' = \vec{\sigma_{x}}$: $\vec{\sigma_{y}}'$ will take $\vec{\sigma_{x}}$. The last two parts guarantee the exclusiveness of the rtoken defined in Property~\ref{prop:1}.

\noindent \textbf{\texttt{copy} Syntax.} The rule for \texttt{copy} syntax is ${ASGNC}$. Since the rvalue is still valid, only one of $\vec{\sigma_{x}}'$ and $\vec{\sigma_{y}}'$ can derive the rtoken if $\vec{\sigma_{x}}$ is holding it. Otherwise, neither of them is the holder. Thus, inference for the new states $\vec{\sigma_x}', \vec{\sigma_y}'$ is required. The constraint consists of two parts: (i) $\vec{\sigma_{y}}=\vec{0}$: the first part is unchanged; (ii) $\{\vec{\sigma_{x}}' = \vec{\sigma_{x}} \land \vec{\sigma_{y}}'=\vec{0}\} \lor \{\vec{\sigma_{y}}' = \vec{\sigma_{x}}\land\vec{\sigma_{x}}'=\vec{0}\}$: the second part guarantees the exclusiveness by distributed to $\vec{\sigma_x}'$ or $\vec{\sigma_y}'$ only. The member who is not the holder must stay zeroed to satisfy Property~\ref{prop:1}. After adding new constraints to update $C'$, the analyzer will remap $x,y$ to the latest state and update the mapping to $M'$.

\begin{table}\small
\caption{The core syntax of Rust MIR.}
\label{table:syntax}
\centering
\resizebox{\linewidth}{!}{
\begin{threeparttable}
\begin{tabular}{ccll}
	\toprule[1pt]
	\textbf{Category}                    & \textbf{Item}            & \multicolumn{1}{c}{\textbf{Syntax}}    & \multicolumn{1}{c}{\textbf{Description}}            \\
	\midrule[1pt]
	\multirow{3}{*}{\textbf{Index}}      & \textbf{BasicBlockId}    & $b \in \mathbb{Z}$                     & Const index of basic block started with 0.          \\
	                                     & \textbf{LocalId}         & $l \in \mathbb{Z}$                     & Const index of local started with 0.            \\
	                                     & \textbf{DefId}           & $f \in \mathbb{Z}$                     & Internal index of function in compiler.             \\
	\midrule[1pt]
	\multirow{4}{*}{\textbf{Variable}}   & \textbf{Local}           & $v \in \{v_0,v_1,...\}$                & Variable and temporary in function.                 \\
	                                     & \textbf{Type}            & $t \in T(P)$                           & Explicit type of local.                         \\
	                                     & \textbf{Place}           & $p=v|v.f|v[i]|*v$                      & Direct, field, index, deference access to the local.        \\
	                                     & \textbf{Operand}         & $o=\ $copy$\ p|$move$\ p$              & The way of create values via loading place.         \\
	\midrule[1pt]
	\multirow{2}{*}{\textbf{Function}}   & \textbf{Statement}       & $s \in STMT(b)$                        & Statements for function body in each block.         \\
	                                     & \textbf{Terminator}      & $j \in TERM(b)$                        & Terminator for function body in each block.         \\
	\midrule[1pt]
	\multirow{2}{*}{\textbf{Liveness}}   & \textbf{StorageLive}     & $live(v)$                              & Mark the beginning of the live range for local.     \\
	                                     & \textbf{StorageDead}     & $dead(v)$                              & Mark the end of the live range for local.           \\
	\midrule[1pt]
    \multirow{5}{*}{\textbf{Statement}}  & \textbf{Assignment}      & $p = o$                                & Yield the operand without changing it.              \\
	                                     & \textbf{Reference}       & $p = \&\ \_\ p$                        & Crate a reference to the place.                     \\
	                                     & \textbf{AddressOf}       & $p = *\ \_\ p$                         & Crate a raw pointer to the place.                   \\
	                                     & \textbf{Casting}         & $p = p \ as \  t$                      & Cast type by using keyword "as".                    \\
	                                     & \textbf{Discriminant}    & $disc(p,c)$                            & Write the discriminant for enum place.              \\
	\midrule[1pt]
	\multirow{4}{*}{\textbf{Terminator}} & \textbf{Goto}            & $goto(b)$                              & Jump to the successor of the current block.           \\
	                                     & \textbf{Return}          & $return$                               & Return from the function.                           \\
	                                     & \textbf{Drop}            & $drop(p,b)$                            & Destruct the place and go to the next block.        \\
	                                     & \textbf{Call}            & $call(f,p,[o],b)$                      & Call function: operand list and return place.       \\

	\bottomrule[1pt]
\end{tabular}
\begin{tablenotes}
	\footnotesize
	\item[1] This table only displays a subset of the syntax relevant to our paper and the category has been modified for analysis. The full syntax can be found in RFC\#1211.
\end{tablenotes}
\end{threeparttable}}
\end{table} 

\subsubsection{Field Primitives} The rtoken is field-sensitive. The arbitrary rtoken is a bit vector, whereas the field representation is a boolean member. Field operations may result in incompatible length issues in bit vectors for some assignments. To address them, we propose two field primitives, \texttt{Ext} and \texttt{Srk}, as listed in Table~\ref{table:pri}.

\begin{table}\small
\caption{The field primitives in leak-free memory model.}
\label{table:pri}
\centering
\resizebox{\linewidth}{!}{
\begin{tabular}{>{\columncolor{mygray}}llll}
	\toprule[1pt]
	\multicolumn{1}{c}{\textbf{Primitive}} & \multicolumn{1}{c}{\textbf{Description}} & \multicolumn{1}{c}{\textbf{Syntax}} & \multicolumn{1}{c}{\textbf{Operation}}\\
	\midrule[1pt]
	\texttt{Ext} ($v.f$)  &  read from field   & $v_1=v.f$         & Sign-extend $v.f$ and meet with init rtoken.\\
	\texttt{Srk} ($v$)    &  write to field    & $v_1.f=v$         & Join $v$ into a bit vector (length = 1).\\
	-                     &  field to field    & $v_1.f=v.f$       & No additional action.\\
	\bottomrule[1pt]
\end{tabular}
}
\end{table} 

\begin{definition}[Extend]\label{def:ext}
	\texttt{Ext} primitive extracts the field of the rvalue. Then it sign-extends the field and meets with the rtoken encoded through the field's type.
\end{definition}

\begin{definition}[Shrink]\label{def:srk}
	\texttt{Srk} primitive joins all the members of the rvalue to generates a one-lengthed bit vector.
\end{definition}

\texttt{Ext} is only used for fields accessed in rvalues, whereas \texttt{Srk} is used for fields accessed in lvalues. Field primitives improve the adaptability of field operations because the lengths of the lvalue and rvalue could not be identical. Since field access is possible at any depth, we set the maximum depth to one and designated other minor accesses untracked.

\subsection{Function Handling}

\subsubsection{Constructor and Destructor}\label{sec:cdtor} 
Each heap item requires a constructor to create the rtoken. Rust provides two kinds of constructors: (i) function constructor: it returns a fully initialized value, such as \texttt{Vec::new}; (ii) local constructor: it declares the object and initializes fields later~\cite{reference}. The initialized field can be used but the object cannot cross the function boundaries before it is fully initialized. All function inputs are treated as fully-initialized values, and the rtokens are constructed before the first statement of the function. The constructor and destructor are defined as follows:

\begin{definition}[Rtoken Constructor]\label{def:ctor}
	A function constructor creates rtoken by type encoding, and a local constructor creates $\vec{0}$ with the same length.
\end{definition}

\begin{definition}[Rtoken Destructor]\label{def:dtor}
	A destructor negates the bit vector of the rtoken constructed by type encoding.
\end{definition}

The local constructor is initialized to zero to facilitate partial initialization, which can be updated by initializing fields later. For destructors like \texttt{drop()}, the rtoken will meet it with a destructor rtoken to achieve deallocation. Since the rtoken destructor negates the bit vector constructed by type encoding, it can destroy the rtoken and keep the leaked field remaining after calling \texttt{drop()}. Therefore, our model is field-sensitive and supports partial initialization, partial deallocation, and partial movement.

\subsubsection{Inter-procedural Refinement} Errors propagation plague inter-procedural analysis based on boolean satisfiability~\cite{moura2008z3}. If it has an incorrect constraint in the callee, the result of constraint solving is unsatisfied (\textit{i.e.,} \texttt{UNSAT}). We observed that the leaks across the semi-automated boundary have specific patterns; thus we use taint analysis to simplify function boundaries; these functions are also used in the bug reporter to suppress false positives. The components are as follows:

\begin{enumerate}[0]
\item[$\bullet$] \noindent The taint source consumes a heap item and returns pointer type or \texttt{ManuallyDrop}.
\item[$\bullet$] \noindent The taint sanitizer consumes a \texttt{ManuallyDrop} or pointer type and returns a heap item.
\end{enumerate}

\begin{figure}[]
\includegraphics[width=0.5\textwidth]{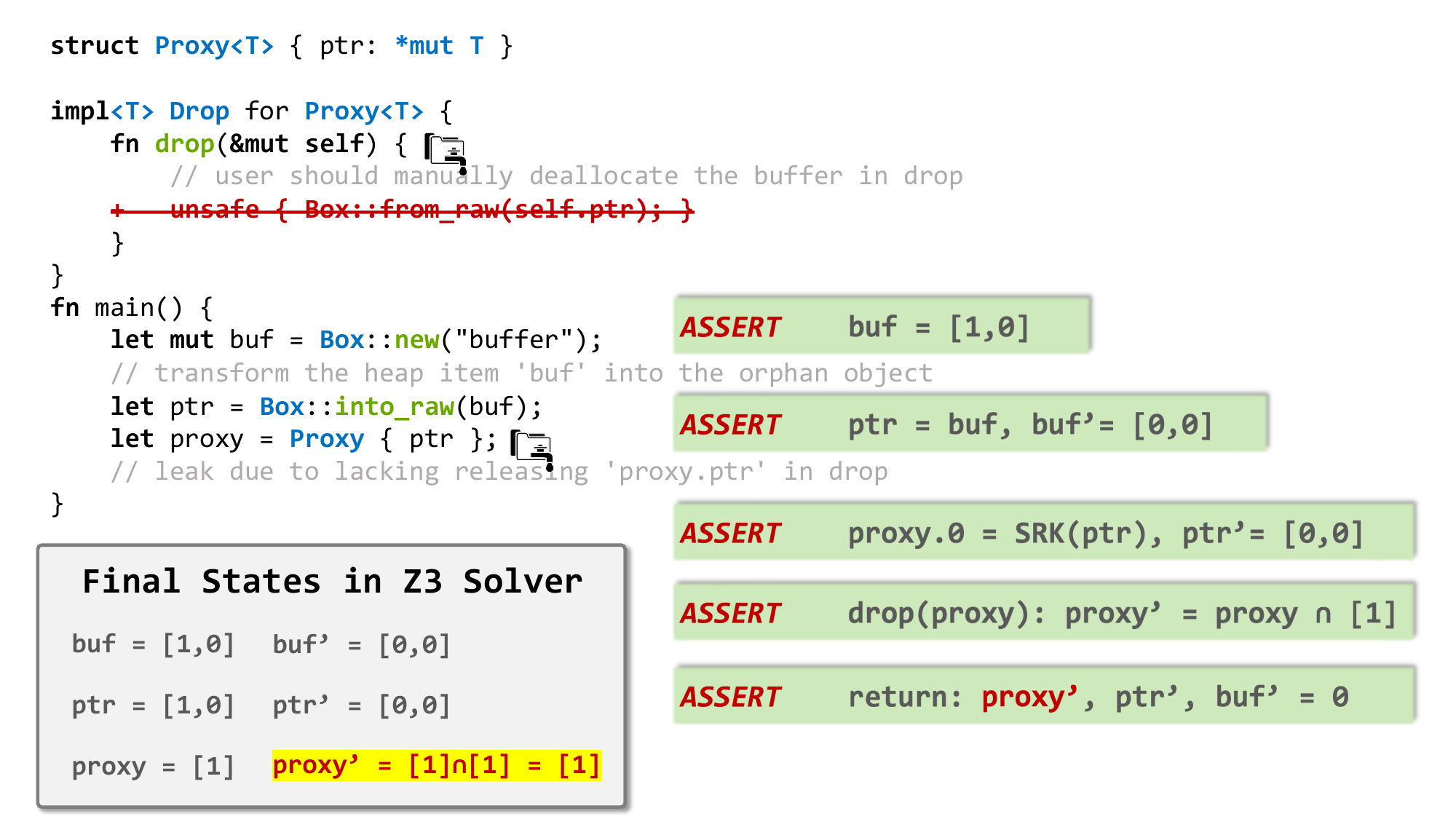}
\caption{Example of the constraints built for the proxy-type leak snippet. Those constraints are manually translated from MIR to the source code to enhance readability.}
\label{fig:example}
\end{figure}

\texttt{Box::leak}~\cite{leak} and \texttt{Box::from\_raw}~\cite{fromraw} are examples of taint sources and sanitizers. The taint source consumes the owner and returns a pointer; the value cannot be freed, or this function will return a dangling pointer. When encountering a taint source, the call site will be recorded as a potential leak candidate if the result of the constraint solver is \texttt{UNSAT}. The taint sanitizer is the opposite that retrieves the owner to activate automated allocation. As for other functions, we assume the object arguments will invoke \texttt{drop()} and other types will not be deallocated by default. The taint analysis can streamline inter-procedural analysis and reduce constraint scales.

\subsection{Example of Leak-free Model}
We illustrate a case of the constraints built upon a proxy-type issue, as shown in Figure~\ref{fig:example}. The green box represents the constraints generated by each statement, while the gray box represents the states of variables when the program analysis is complete. \texttt{Box::new} is a function constructor that creates the heap item \texttt{Box<\&str>} and generates rtoken $[1,0]$ held by \texttt{buf}. \texttt{Box::into\_raw} is a taint source, that transfers the rtoken to \texttt{ptr}. This call site is recorded by the bug reporter. \texttt{ptr} is then assigned to the field of \texttt{proxy}, which is created by the local constructor. Since the length of \texttt{Proxy} $1$ does not match \texttt{ptr} ($2$), the field access requires \texttt{SRK} primitive. As for \texttt{drop()}, the destructor rtoken ($[1]$) is generated by negating the bit vector $[0]$ of the constructor rtoken for \texttt{Proxy}. We intersect this destructor rtoken with the current state to simulate deallocation. When encountering a return, it asserts that the latest states of variables are zeroed (\textit{i.e.,} \texttt{buf'}, \texttt{ptr'}, and \texttt{proxy'}), indicating that all of them are released. If all constraints above are satisfied, the function is confirmed to be leak-free. However, due to the destructor of \texttt{Proxy} cannot delete the field, the result of constraint solving is \texttt{UNSAT}, and the recorded call site will be emitted to the user.

\begin{table*}[]
\caption{Effectiveness evaluation results of \textsc{rCanary} versus \textsc{FFIChecker}, \textsc{Saber}, \textsc{ASAN}, \textsc{Miri}, and \textsc{libFuzzer} on nine package benchmarks. These bugs are non-trivial from their pull requests (PR); the types of bugs include orphan object (OO) and proxy type (PT). The reports are classified into true positives (TF) and false positives (FP).}
\label{table:RQ1}
\resizebox{\linewidth}{!}{
\begin{threeparttable}
\begin{tabular}{lcccccccc>{\columncolor{mygray}}ccccccc}
    \toprule[1pt]
	\multicolumn{7}{c}{\textbf{Package}} & \multicolumn{2}{c}{\textbf{Issue}} & \multicolumn{2}{c}{\textbf{\textsc{rCanary}}} & \multicolumn{1}{c}{\textbf{\textsc{FFIChecker}}} & \multicolumn{1}{c}{\textbf{\textsc{Saber}}} & \multicolumn{1}{c}{\textbf{\textsc{ASAN}}} & \multicolumn{1}{c}{\textbf{\textsc{Miri}}} & \multicolumn{1}{c}{\textbf{\textsc{libFuzzer}}} \\
	\cmidrule(r){1-7}
	\cmidrule(r){8-9}
	\cmidrule(r){10-11}
	\cmidrule(r){12-12}
	\cmidrule(r){13-13}
	\cmidrule(r){14-14}
	\cmidrule(r){15-15}
	\cmidrule(r){16-16}
	Name & Functions & LoC & Tests & AdtDef & Ty & Taint & PR & Pattern & TP & FP & TP/FP & TP/FP & TP/FP & TP/FP & TP/FP \\
	\midrule[1pt]
	napi-rs        & 1428  & 20.9k   & 14   & 724    & 2408  & 86    & \#1230  & PT & 1+18  & 4  & 0/0  & 0/5   & FAIL  & FAIL   & FAIL\\ 
	%
	rust-rocksdb   & 3295  & 30.2k   & 43   & 503    & 1593  & 17    & \#658   & OO & 1+16  & 0  & 1/0  & 0/8   & 0/13  & FAIL   & 1/0\\ 
	arrow-rs       & 6164  & 149.6k  & 1202  & 1293  & 7452  & 21    & \#1878  & OO & 1+0   & 0  & FAIL & FAIL  & 0/1   & 2/8    & FAIL\\ 
	arma-rs        & 179   & 2.2k    & 66    & 243   & 858   & 3     & \#22    & OO & 1+0   & 0  & 0/0  & 0/6   & 0/1   & 0/0    & 0/0\\ 
	ruffle         & 6474  & 135.8k  & 915   & 11301 & 67092 & 5     & \#6528  & PT & 1+0   & 0  & 0/0  & 0/12  & FAIL  & FAIL   & FAIL\\ 
	flaco          & 22    & 0.4k    & 0     & 258   & 729   & 14    & \#12    & PT & 2+0   & 0  & FAIL & 0/3   & 0/0   & 0/0    & FAIL\\ 
	pprof-rs       & 110   & 2.3k    & 9     & 162    & 508  & 3     & \#84    & PT & 1+0   & 0  & 1/0  & 0/2   & FAIL  & 0/0    & 1/0\\ 
	rowan          & 406   & 4.4k    & 5     & 118    & 442  & 10    & \#112   & PT & 1+0   & 3  & 1/0  & 0/7   & 0/0   & 0/0    & 0/0\\ 
	Fornjot        & 688   & 11.8k   & 40    & 304    & 13141 & 2   & \#646  & PT & 1+0   & 0  & 0/0  & FAIL  & 0/0   & 0/0    & 1/0\\ 
    \bottomrule[1pt]
\end{tabular}
\begin{tablenotes}\footnotesize
	\item Failed: \textsc{FFIChecker}: internal errors (self abort due to panic); \textsc{Saber}: the minimum Rust version supported (LLVM14) conflicts with the maximum \textsc{SVF} support (LLVM13); \textsc{ASAN}: cannot integrate \textsc{LeakSanitizer} into cargo tests; \textsc{Miri}: does not support to test \texttt{extern} FFI calls; \textsc{libFuzzer}: cannot generate fuzz targets through \texttt{cargo-fuzz}.
\end{tablenotes}
\end{threeparttable}
}
\end{table*}

\begin{table*}[]
\caption{Real-world vulnerability evaluation results of \textsc{rCanary} with the detailed issue description. After the manual check, these vulnerabilities are classified into orphan-object (OO) and proxy-type (PT), along with their leak scenarios.}
\label{table:new_bug}
\resizebox{\linewidth}{!}{
\begin{threeparttable}
\begin{tabular}{lccc>{\columncolor{mygray}}ccl}
    \toprule[1pt]
	\textbf{Crate} & \textbf{Functions} & \textbf{LoC} & \textbf{Location} & \textbf{Pattern} & \textbf{Summary}  & \textbf{Description}\\
	\midrule[1pt]
	 basic\_dsp \underline{\#52}    & 1998   & 30.3k   & mod.rs         & PT     & LeakedDropImpl       & Struct \texttt{VectorBox$<$B,T$>$} leaks the field \textit{argument} in its \texttt{Drop} \texttt{impl}. \\
	 symsynd \underline{\#15}       & 35     & 0.5k    & cabi.rs          & PT   & LeakedDropImpl       & Field \textit{message} stored a boxed slice in struct \texttt{CError} should be freed in \texttt{Drop} \texttt{impl}.\\
	 rustfx \underline{\#3}         & 263    & 4.8k    & core.rs*         & PT   & LeakedDropImpl       & Field \textit{host} in struct \texttt{OfxHost} and \textit{String} in enum \texttt{PropertyValue}.\\
	 \midrule[1pt]
	 signal-hook \underline{\#150}  & 234    & 5.4k    & raw.rs          & OO   & Overwriting  & Calling \texttt{init()} multiple times will leak \texttt{AtomicPtr} in \textit{slot} and trigger panicking. \\
	 rust-vst2 \underline{\#42}     & 207    & 4.1k    & host.rs         & OO   & Overwriting  & Global \texttt{LOAD\_POINTER} in \texttt{call\_main()} will leak \texttt{host} for multiple assignments. \\
	 synthir \underline{\#1}        & 489    & 7.9k    & emulator.rs     & OO   & Overwriting  & \texttt{MASKS\_ALL} and \texttt{MASKS\_BIT} may leak the pointed \texttt{Vec$<$BigUint$>$}.\\
	 relibc (redox-os) \underline{\#180}  & 1709   & 32.0k   & mod.rs*   & OO   & Overwriting   & Calling \texttt{init()} will overwrite \texttt{PTHREAD\_SELF}; \textit{packet\_data\_ptr} may be leaked.\\
	 tinyprof \underline{\#1}       & 26     & 0.5k    & profiler.rs     & OO   & Overwriting  & Assign by dereferencing to \texttt{PROFILER\_STATE\_SENDER} will leak the front chunck.\\
	 teaclave-sgx \underline{\#441} & 7558   & 147.4k  & func.rs*        & OO   & Overwriting   & \textit{session\_ptr} and \textit{p\_ret\_ql\_config} may be leaked if overwrites to it.\\
	 tor\_patches \underline{\#1}  & 288    & 4.1k    & tor\_log.rs     & OO   & Overwriting  & Rewriting to \texttt{LAST\_LOGGED\_FUNCTION} and \texttt{LAST\_LOGGED\_MESSAGE} will leak box.\\
	 \midrule[1pt]
	 log \underline{\#314}         & 438    & 5.0k    & lib.rs           & OO   & Finalization                   & \texttt{set\_boxed\_logger} leaks a boxed \texttt{Log} to \texttt{LOGGER} along with potential spin loop. \\
	 tracing \underline{\#1517}    & 3375   & 61.0k   & dispatch.rs      & OO   & Finalization  & Intentional leak towards a global \texttt{GLOBAL\_DISPATCH}, the relevant issue in \#1517. \\ 
	 rust-appveyor \underline{\#1} & 43753  & 733.4k  & thread\_local.rs & OO   & Finalization   & \texttt{KEYS} and \texttt{LOCALS} need finalization function to clean up.\\
	 ServoWsgi         & 11606  & 236.2k  & opts.rs          & OO   & Lazy   & \texttt{DEFAULT\_OPTIONS} needs finalization to clean up, especially for multi threads exit.\\
	 rux                & 585    & 9.9k    & lazy.rs          & OO   & Lazy     & The current lazy static uses the \textit{'static} that cannot be freed in multi threads.\\
	 next\_space\_coop  & 20477  & 317.1k  & lazy.rs          & OO   & Lazy     & The current lazy static uses the \textit{'static} that cannot be freed in multi threads.\\
	\midrule[1pt]
	 rio \underline{\#52}  & 95     & 3.0k    & lazy.rs          & OO   & PanicPath & If the assertion goes into unwinding, the \textit{value} will be leaked inside a panic.\\
	 sled \underline{\#1458} & 1350   & 32.8k   & lazy.rs          & OO   & PanicPath  & If the assertion goes into unwinding, the \textit{value} will be leaked inside a panic.\\
    \bottomrule[1pt]
\end{tabular}
\begin{tablenotes}\footnotesize
	\item The item having \texttt{*} indicates vulnerabilities located in multiple files. Additionally, we did not open an issue specifically for scenario \texttt{Lazy} because it is a common design flaw in lazy\_static implementations.
\end{tablenotes}
\end{threeparttable}
}
\end{table*}


\section{Evaluation}
We built a prototype of \textsc{rCanary} as an external component for \texttt{Cargo} and \texttt{rustc} and added around 8k lines of Rust code. The analyzer is implemented specifically upon the rustc-v1.62 toolchain. We have conducted comprehensive evaluations of \textsc{rCanary} using real-world Rust programs. Our evaluation aims to answer the following research questions (RQs):

\begin{enumerate}[0]
\item[$\bullet$] \noindent \textbf{RQ1.} How effective is \textsc{rCanary} at detecting existing leak bugs? 
\item[$\bullet$] \noindent \textbf{RQ2.} How capable is \textsc{rCanary} of finding real-world leak vulnerabilities?
\item[$\bullet$] \noindent \textbf{RQ3.} How efficient is \textsc{rCanary} in analyzing Rust ecosystem?
\item[$\bullet$] \noindent \textbf{RQ4.} How is \textsc{rCanary} compared to other approaches?
\end{enumerate}

\noindent\textbf{Experimental Setup.} The following experiments were conducted on a machine with a 2GHz Intel CPU and 32GB of RAM running 64-bit Ubuntu LTS 22.04.

\subsection{Effectiveness Evaluation (RQ1, RQ4)}\label{sec:effectiveness}

\subsubsection{Benchmark Generation} Since \textsc{rCanary} is the first static work aimed at memory leaks across a semi-automated memory management model in Rust, we do not have a benchmark that previously existed to evaluate. Therefore, we devised an objective benchmark containing nine crates with fixed pull requests of leak issues as the ground truth for evaluation; this benchmark can also be utilized in the subsequent study. To evaluate the effectiveness, several leak analyzers are applied to these crates.

The benchmark generation rigorously follows the procedure below. We retrieve results searching from GitHub using the keyword \textit{\textbf{Memory Leak}}, restricting the language to Rust, and concentrating exclusively on the \textbf{Issue} and \textbf{Rull Request} sections of repositories, sorted by Github's default order. We filter the top 30 pages and exclude the following irrelevant items:

\begin{enumerate}[0]
\item[$\bullet$] \noindent Leaks caused by reference cycles.
\item[$\bullet$] \noindent Leaks caused solely by manual memory management.
\item[$\bullet$] \noindent Leaks caused by outside FFI functions.
\item[$\bullet$] \noindent Leaks referred to sensitive information leaking (\textit{e.g.,} private key leaks).
\item[$\bullet$] \noindent Repositories used for language teaching or learning.
\end{enumerate}

After applying the above filters, the authors audit the buggy snippet and confirm that leaks were caused across the semi-automated boundaries, matching our bug patterns. Finally, nine repositories are retained, as listed in Table~\ref{table:RQ1}.

\subsubsection{Comparing Tools} We compare our tool to the static bug detector \textsc{FFIChecker}~\cite{li2022detecting} for Rust-C FFI, the static leak analyzer \textsc{Saber} (from \textsc{SVF}~\cite{sui2016svf}), the runtime leak detector \textsc{LeakSanitizer} (from Google \textsc{AddressSanitizer}~\cite{serebryany2012addresssanitizer}), the MIR interpreter \textsc{Miri}~\cite{olson2016miri,jung2019stacked}, and the fuzzing tool \textsc{\textsc{libFuzzer}} (from LLVM)~\cite{serebryany2016continuous}. Because numerous static leak detectors do not support Rust projects, we can only compare LLVM-based approaches~\cite{lattner2004llvm}.

\textsc{FFIChecker} is built on a specific Rust version and only supports Rust-C FFI. \textsc{Saber} is a component of the source code analyzer \textsc{SVF} that employs inter-procedural sparse value-flow analysis to detect leaks for LLVM-based languages. Due to compatibility needs for LLVM, we modified \textsc{Saber} to conform to the allocators in Rust 1.57. Other methods are based on dynamic analysis. \textsc{LeakSanitizer} is a memory leak detector, but an executable file or unit test is required to expose vulnerabilities. Similarly, \textsc{Miri} requires the attribute \texttt{\#[text]} to annotate the unit tests. As for \textsc{libFuzzer}, we employ \texttt{cargo-fuzz} to manually activate fuzzing to buggy functions after generating fuzz targets. \textsc{LeakSanitizer}, \textsc{Miri}, and \textsc{libFuzzer} are components in Rust, so we invoke them via the Rust 1.62 toolchain.

\subsubsection{Bug Report} Our analyzer integrated a filter into the bug reporter to refine the output. After analyzing bug reports in the experiments, we identified the following factors that can be the root cause of irrelevant outputs: (i) \textbf{Callee Propagation}: incorrect constraints or leaks in the callees will affect callers. (ii) \textbf{Extern Function}: many extern functions intentionally leak resources in the return value for FFI use. Thus, we devise a filter integrated into the bug reporter to filter the irrelevant items, although they may trigger memory leaks. The filter only reports the taint source in the top entry and ignores the return value of extern functions if any leak exists.

\subsubsection{Results} \textsc{rCanary} can recall all related issues with acceptable false positives, as shown in Table~\ref{table:RQ1}. The number behind \texttt{+} indicates that the developers should manually check the FFI functions called by Rust that can free the orphan objects because Rust would not deallocate them. \textsc{FFIChecker} can detect three of them, but it only supported Rust-C FFI and LLVM-IR lost the language information. \textsc{Saber} fails to recall any issues because its report provides no valuable information for users to locate the bug. Like \textsc{FFIChecker}, the primary cause is that LLVM-IR lost high-level programming language features. \textsc{LeakSanitizer}, \textsc{Miri}, and \textsc{libFuzzer} cannot evaluate generic functions and highly depend on the coverage of sufficient unit tests for leak functions. Furthermore, \textsc{Miri} is insufficient for functions that invoke FFI calls. Although \textsc{rCanary} supports functions with the \texttt{extern} keyword, it still cannot cross the FFI boundary. Since compared tools are limited in leak detection for Rust semi-automated boundaries, they have better pattern or language support than \textsc{rCanary}.

\subsection{Real-world Vulnerability Evaluation (RQ2)}\label{sec:new_bug}
More than 1,200 real-world Rust crates were downloaded and analyzed from crates.io and GitHub. All of these packages are evaluated by using a central script. The analyzer takes each local function as a distinct entry with a new context for each project and checks the satisfiability of the leak-free memory model to detect potential leaks in their function bodies.

Among the evaluated packages, 3.0\% have conflicted Rust version after switching toolchain, 11.6\% have an unsupported virtual \texttt{Cargo} workspace, and 85.4\% have been effectively analyzed. After diagnostics are sent to the bug reporter, we manually check the source code and categorize them based on leak patterns and real-world scenarios. The report outputs the vulnerabilities with related functions and further pinpoints the detailed bug locations. 

We listed 19 crates with memory leaks, and the detailed descriptions are listed in Table~\ref{table:new_bug}. Those vulnerabilities are non-trivial with different scenarios. For example, the orphan-object leak in the popular crate \texttt{signal-hook} from crates.io exposes that repeated assignment to a \texttt{static} \texttt{mut} pointer can cause leaks in the local thread. The proxy-type vulnerabilities in \texttt{basic\_dsp} have generic-type parameters invisible for dynamic tests without mono-morphization. We observe that the reassignment of the static pointer is error-prone to memory leaks, which will be discussed in the case study.

\subsection{Performance Evaluation (RQ3)}\label{sec:efficiency}

\begin{figure}[]
\includegraphics[width=0.46\textwidth]{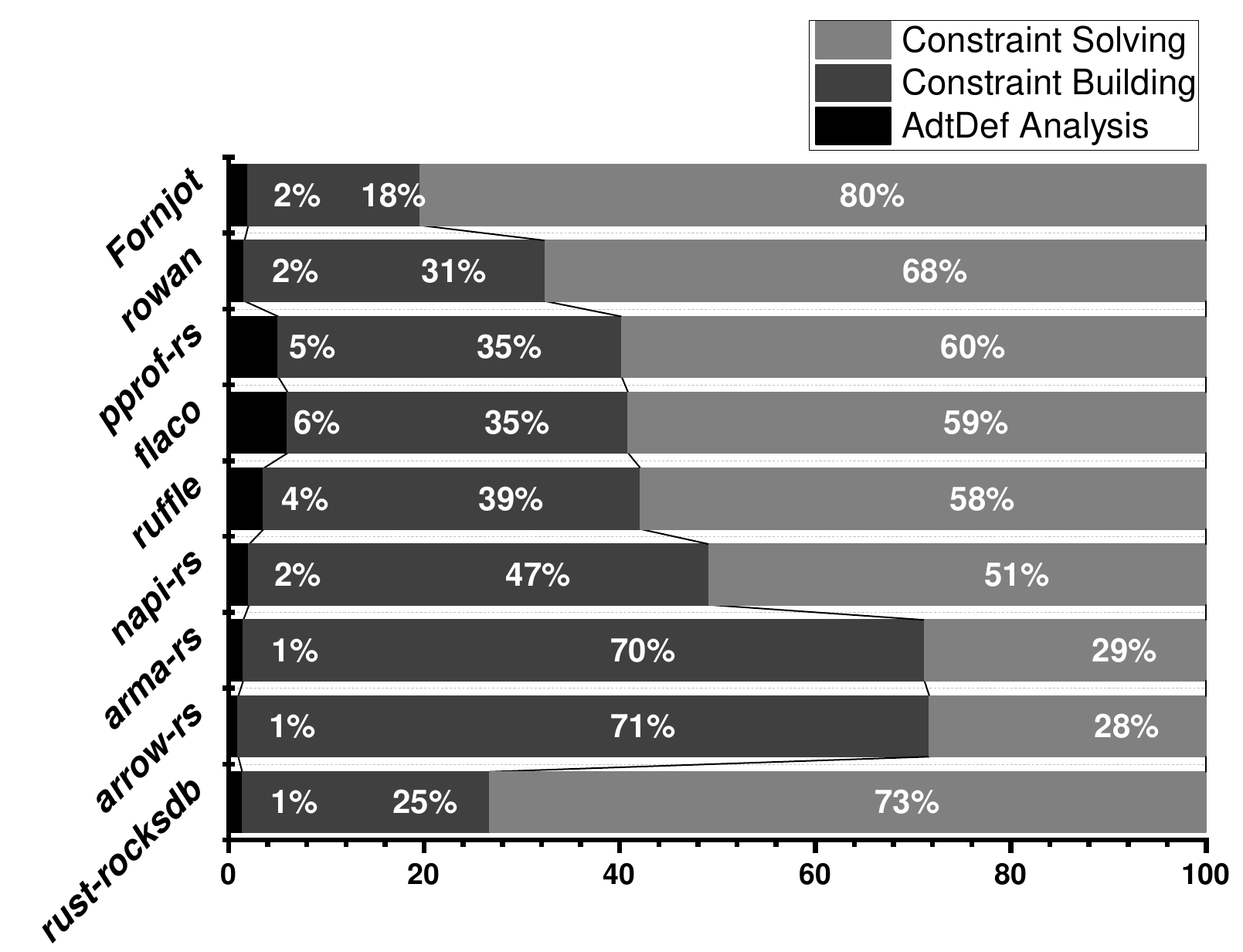}
\caption{The time distribution results on the benchmark. For the virtual workspace, we individually test each crate inside and calculate the average percentage distribution in three phases.}
\label{fig:efficiency}
\end{figure}

We expect \textsc{rCanary} to be an efficient static scanner for the Rust ecosystem. It should not impose significant overhead on the target. At the ecosystem scale, \textsc{rCanary} took about 96 minutes to scan 1.2k real-world Rust packages. It took 8.4 seconds on average to analyze each package. Since we use incremental compilation, it does contain the overhead of the compilation elapsed. Furthermore, there was no aborting or panicking while running the analyzer on evaluated crates. That indicates \textsc{rCanary} is promising to be a general, reliable, and fast source-code leak scanner for the Rust ecosystem.

We examine the time distribution of the overhead among three phases toward our benchmarks in Figure~\ref{fig:efficiency}. All experiments were repeated five times. The cost can be divided into AdtDef analysis, constraint building, and constraint solving. The median distributions are AdtDef analysis at 5\%, constraint building at 35\%, and constraint solving at 59\%. The algorithm of AdtDef analysis is efficient, and the constraint-related phases cost most of the time (more than 95\%). The cost of constraint building and constraint solving does not illustrate explicit relevance because the front contains type encoding which suffers from macros expansion due to generating massive types that may drag down the speed (\textit{e.g.,} the types count in \texttt{ruffle}). As our algorithm does not support pointer arithmetic, the solver's performance might be improved if such operations are commonly used as in crate \texttt{arma-rs}. 
\subsection{Case Study}
As globals are represented as normal variables in MIR, we did not initially consider peculiarities when designing \textsc{rCanary}. However, real-world vulnerabilities have revealed some issues. Many programmers use \texttt{static} \texttt{mut} values to store global data on the heap chunk, but numeral implementations are unsound, including:

\begin{enumerate}[0]
\item[$\bullet$] \noindent Unsound initialized functions that can overwrite the previously stored value without deallocation, potentially leading to denial-of-service attacks or memory exhaustion.
\item[$\bullet$] \noindent Static values that are thread-unsafe or non-finalizing which cannot be deallocated when threads exit, including \texttt{static} \texttt{mut} and \texttt{lazy} values.
\item[$\bullet$] \noindent Incapability to deal with the poisoned static values if the first initialization panics.
\end{enumerate}

\begin{table}\small
\caption{The attributes for the real-world leak scenarios.}
\label{table:attr}
\centering
\resizebox{\linewidth}{!}{
\begin{tabular}{ccccc}
	\toprule[1pt]
	\textbf{Scenario} & \textbf{Panic Only} & \textbf{Invalid Use} & \textbf{Work Around} & \textbf{Memory Safe}\\
	\midrule[1pt]
	LeakedDropImpl  &  $\times$      &  $\checkmark$   & $\checkmark$   & $\times$\\
	Overwriting     &  $\times$      &  $\times$       & $\checkmark$   & $\checkmark$\\
	Finalization    &  $\times$      &  $\times$       & $\times$       & $\checkmark$\\
	Lazy            &  $\times$      &  $\checkmark$   & $\times$       & $\checkmark$\\
	PanicPath       &  $\checkmark$  &  $\checkmark$   & $\checkmark$   & $\checkmark$\\
	\bottomrule[1pt]
\end{tabular}
}
\end{table}

We summarize the attributes for real-world leak scenarios in Table~\ref{table:attr}. All of them may lead to performance degradation, but only \textit{panicpath} is caused by panic. \textit{Lazy} and \textit{finalization} are valid uses, where leaking thread-local values that are not harmless if only one thread. \textit{Leakeddropimpl}, \textit{overwriting}, and \textit{panicpath} have workaround functions that can be repeatedly invoked to consume memory quickly, potentially leading to memory exhaustion and denial of service. Among them, \textit{overwriting} has the lowest cost. As for \textit{Leakeddropimpl}, even if the \texttt{Drop} is fixed (Listing~\ref{list:me_pt}), it is not memory-safe if users assign fields from the pointers that are not orphan objects. Therefore, the current design has significant security risks.

\subsection{Discussion}
Like most static analyzers, \textsc{rCanary} is neither sound nor complete. The heap-item unit hinges on \texttt{PhantomData}. It conventionally covers collection types in the standard library but may cause false positives for some user-defined types. The field primitives cannot eliminate false negatives introduced by restricting the depth of the field access, which can be refined in future work. Furthermore, our model is insufficient if users store orphan objects in arrays or containers (\textit{e.g.,} \texttt{Vec} and \texttt{HashMap}), and it does not support pointer arithmetic either.

We also discovered that the boundary between automated and manual deallocation is unclear in FFI scenarios. To prevent leaks across the semi-automated boundary, we have the insight that it is necessary to design a wrapper like \texttt{ManuallyDrop} with a set of APIs. It needs to downcast the owner for pointer types and use an explicit annotation for external functions to check the double-free or leak issues in FFI via static analysis.

\section{Related Work}

\noindent \textbf{Ownership model.} \textsc{rCanary} adds formal rules to the Rust ownership model. Leak-free memory model shares many similarities with ownership syntax~\cite{heine2003practical,clarke2001object}, and C++ smart pointers~\cite{edelson1992smart}. It resembles linear types~\cite{wadler1990linear,girard1995linear} to support Rust syntax features such as ownership movement~\cite{pierce2004advanced}. Clarke et al.~\cite{heine2003practical} proposed a strict ownership model named ownership type that asserts the owning relation between objects~\cite{noble1998flexible,clarke1998ownership}. Clouseau~\cite{heine2003practical} adds the features of ownership transfer, arbitrary aliases, and ownership invariant at function boundaries to enhance scalability, which is one of the foundations of our model. Although some language extensions are proposed to allow read-only aliases~\cite{boyland2001alias,walker2000alias}, the strict definition does not require right-hand-side pointers~\cite{wadler1990linear}. Our model allows pointers to access objects, but the rtoken holder must be exclusive~\cite{boyapati2002ownership,clarke2002ownership}. We extend the proprietor to all pointer types and provide an explicit boundary between stack-only and heap-allocated data for Rust types.

\noindent \textbf{Static leak detection.} Numerous static analysis studies focus on leak detection. \textsc{FFIChecker}~\cite{li2022detecting} is designed for Rust-C FFI based on LLVM IR. \textsc{Rupta}~\cite{li2024context} is a flow-insensitive but context-sensitive point-to alias analysis model designed for Rust MIR, with the potential to be used for leak detection in the future. Saturn~\cite{xie2005context} is based on boolean satisfiability, which is context- and path-sensitive. \textsc{rCanary}, unlike Saturn, is a path-insensitive but flow-sensitive checker. Clouseau~\cite{heine2003practical} is a flow- and context-sensitive ownership model, but our work has a field-sensitive data abstraction and provides features like partial initialization. Some studies employ symbolic execution~\cite{king1976symbolic} and abstract interpretation~\cite{cousot1977abstract,cousot1979systematic} to detect leaks for C/C++ programs such as KLEE~\cite{cadar2008klee}, Clang~\cite{lattner2008llvm}, and Sparrow~\cite{jung2008practical}. Also, the full-sparse value-ﬂow analysis~\cite{sui2014detecting,sui2016svf,hardekopf2011flow} monitors value flow and define-use chains via top-level and address-taken pointers to detect leaks. Like SVF, semi-sparse ﬂow-sensitive analysis~\cite{cherem2007practical,hardekopf2009semi} can identify define-use chains by explicitly placing top-level pointers in SSA form.

\section{Conclusion}
Semi-automated memory management model cannot prevent memory leaks in Rust. In this paper, we studied the root cause of memory leaks caused by this model and summarized bug patterns. We presented \textsc{rCanary}, a static, non-intrusive, and fully automated model checker based on boolean satisﬁability to detect them. We implemented this tool on top of Rust MIR and evaluated it through real-world Rust projects. \textsc{rCanary} can efficiently recall our benchmarks and find 19 crates with memory leaks with high efficiency.

\makeatletter
\def\endthebibliography{%
\def\@noitemerr{\@latex@warning{Empty `thebibliography' environment}}%
  \endlist
}
\makeatother
\bibliographystyle{IEEEtran}
\bibliography{rcanary}

\end{document}